%% file: main.tex
\documentclass{article}
\usepackage{PRIMEarxiv}

\usepackage{hyperref}
\makeatletter
\def\UrlAlphabet{%
      \do\a\do\b\do\c\do\d\do\e\do\f\do\g\do\h\do\i\do\j%
      \do\k\do\l\do\m\do\n\do\o\do\p\do\q\do\r\do\s\do\t%
      \do\u\do\v\do\w\do\x\do\y\do\z\do\A\do\B\do\C\do\D%
      \do\E\do\F\do\G\do\H\do\I\do\J\do\K\do\L\do\M\do\N%
      \do\O\do\P\do\Q\do\R\do\S\do\T\do\U\do\V\do\W\do\X%
      \do\Y\do\Z}
\def\UrlDigits{\do\1\do\2\do\3\do\4\do\5\do\6\do\7\do\8\do\9\do\0}
\g@addto@macro{\UrlBreaks}{\UrlOrds}
\g@addto@macro{\UrlBreaks}{\UrlAlphabet}
\g@addto@macro{\UrlBreaks}{\UrlDigits}
\makeatother

\usepackage[utf8]{inputenc} 
\usepackage[T1]{fontenc}    
\usepackage{booktabs}       
\usepackage{amsfonts}       
\usepackage{nicefrac}       
\usepackage{microtype}      
\usepackage{lipsum}
\usepackage{fancyhdr}       
\usepackage{graphicx}       
\graphicspath{{media/}}     

\input{cmd.tex}
\title{COX: CUDA on X86 by Exposing Warp-Level Functions to CPUs} 

\author{
  Ruobing Han \\
  Georgia Institute of Technology \\
  USA \\
  \texttt{hanruobing@gatech.edu} \\
     \And
  Jaewon Lee \\
  Georgia Institute of Technology \\
  USA \\
  \texttt{jaewon.lee@gatech.edu} \\
   \And
  Jaewoong Sim \\
  Seoul National University \\
  Korea \\
  \texttt{jaewoong@snu.ac.kr} \\
  \And
 Hyesoon Kim \\
  Georgia Institute of Technology \\
  USA \\
  \texttt{hyesoon@cc.gatech.edu} \\
}

\begin{document}

\maketitle

\begin{abstract}
As CUDA programs become the de facto program among data parallel applications such as high-performance computing or machine learning applications, running CUDA on other platforms has been a compelling option. Although several efforts have attempted to support CUDA on other than NVIDIA GPU devices, due to extra steps in the translation, the support is always behind a few years from supporting CUDA's latest features. The examples are DPC, Hipfy, where CUDA source code have to be translated to their native supporting language and then they are supported. In particular, the new CUDA programming model exposes the warp concept in the programming language, which greatly changes the way the CUDA code should be mapped to CPU programs. In this paper, hierarchical collapsing that \emph{correctly} supports CUDA warp-level functions on CPUs is proposed. Based on hierarchical collapsing, a framework, COX, is developed that allows CUDA programs with the latest features to be executed efficiently on CPU platforms. COX consists of a compiler IR transformation (new LLVM pass) and a runtime system to execute the transformed programs on CPU devices. COX can support the most recent CUDA features, and the application coverage is much higher (90\%) than for previous frameworks (68\%) with comparable performance. We also show that the warp-level functions in CUDA can be efficiently executed by utilizing CPU SIMD (AVX) instructions. 
\end{abstract}


\section{Introduction}
The high-performance computing power in GPUs has developed a strong software eco-system based on GPU programs.
Although there are other choices for GPU programming such as HIP~\cite{hip}, OpenMP and DPC~\cite{dpct}, in recent years, CUDA is still in the dominant place. In the realm of Deep Learning, both of the two most popular frameworks, Pytorch and TensorFlow, support only CUDA for GPU backend \cite{asaduzzaman2021impact}. In the multimedia realm, for the video editing applications, as lists in \cite{GPU-roundup}, CUDA is compatibility in 14 of the 17 applications, while OpenCL is only supported by 9 of them. 

Unfortunately, despite the popularity of CUDA programs, NVIDIA GPUs are the main hardware platforms to run them. Although there have been several efforts to run CUDA on non-NVIDIA GPUs, they are lack supporting newer CUDA features. 
There are mainly two challenges to run CUDA on other platforms are. The first is to convert Single Program Multiple Data (SPMD) programs to programs for non-SPMD friendly architectures. In the SPMD programming model, the same kernel is executed by many threads at runtime, and the GPU is built on for through-put-oriented architectures by supporting many threads (or warps). However, other architectures often do not have that many threads, so the CUDA programs need to be converted to a fewer number of threads efficiently. The second problem is a continuous support for still evolving programming models like CUDA. To address the second problem, we utilize the open-source compiler frame LLVM as much as possible. In this paper, we tackle the first problem: Supporting CUDA with fewer number threads, which is an essential component for running CUDA on X86, ARM, or Intel-GPU, which has fewer hardware threads than NVIDIA GPUs.

\begin{figure*}[ht]
    \centering
	\begin{subfigure}[t]{0.33\textwidth}
			\includegraphics[width=\textwidth,trim=4 4 4 4,clip]{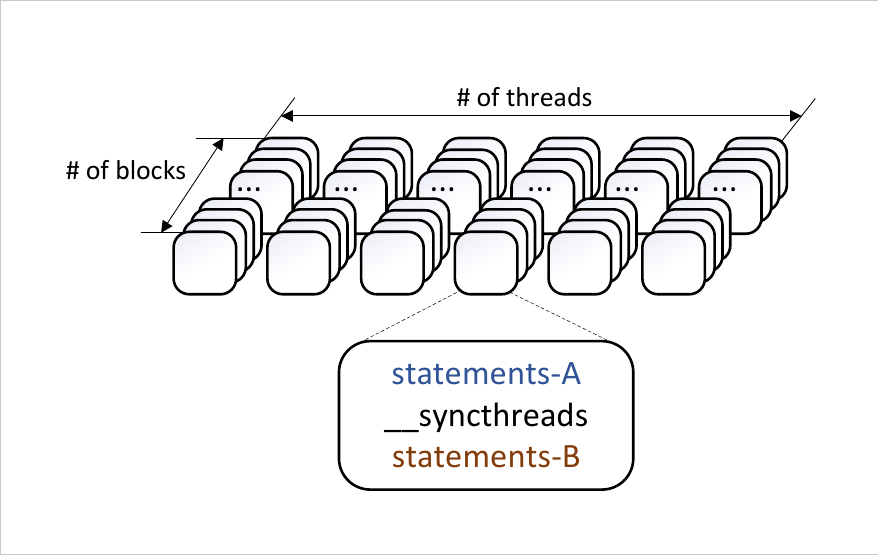}
		\caption{GPU SPMD}		
	\end{subfigure}
    \begin{subfigure}[t]{0.27\textwidth}
		\includegraphics[width=\textwidth,trim=4 4 4 4,clip]{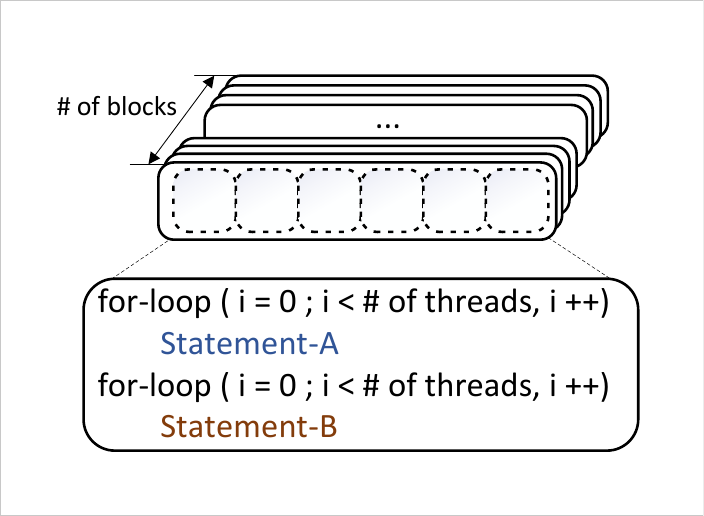}
		\caption{Output of flat collapsing}			
	\end{subfigure}
    \begin{subfigure}[t]{0.33\textwidth}
		\includegraphics[width=\textwidth,trim=4 4 4 4,clip]{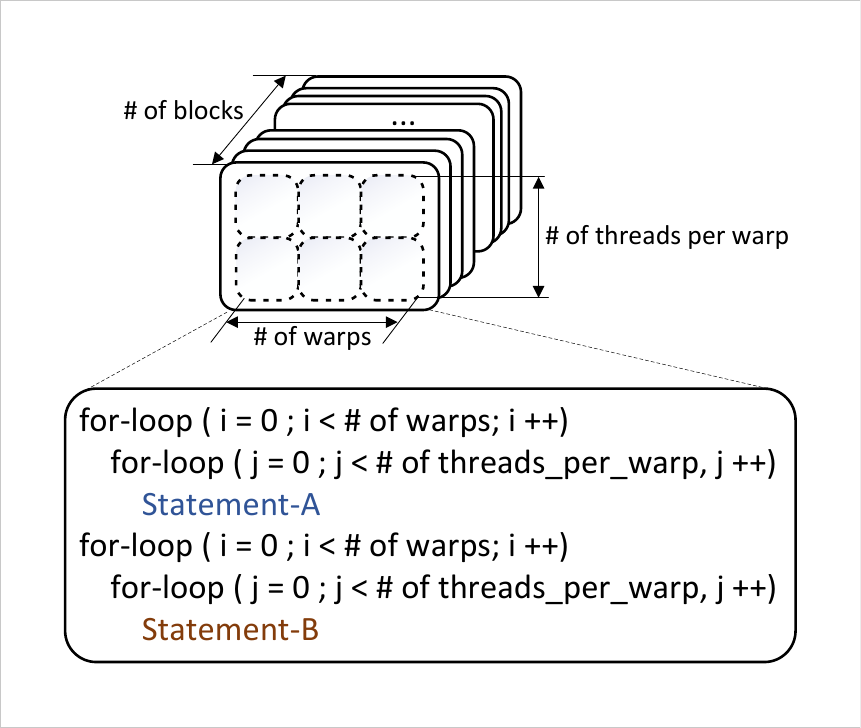}
		\caption{Output of hierarchical collapsing}			
	\end{subfigure}
    \caption{The programming model for input CUDA SPMD and output CPU programs by flat collapsing and hierarchical collapsing}
    \label{fig:SPMD}
\end{figure*}

Several projects aim to support this transformation: running GPU programs on CPUs~\cite{stratton2008mcuda, diamos2010ocelot,jaaskelainen2015pocl,stratton2010efficient,hipify,dpct,intel-opencl,PGI,karrenberg2012improving,diamos2010dynamic,CudaEmulator,chen2018enabling,gummaraju2010twin}. While a few projects focus on narrowing the gap between SPMD and MPMD by adding a hardware extension\cite{chen2018enabling} or adding system-level support for faster context switching\cite{gummaraju2010twin}, most projects try to do compiler-level transformation to translate GPU functions to be suitable for CPUs. These projects use the same granularity of transformation: a CPU thread is responsible for executing all the threads in a CUDA block (or OpenCL work-group). The CUDA-block-to-CPU-thread mapping is optimal based on three observations: 1) it has a fewer CPU threads compared with CUDA-thread-to-CPU-thread mapping, which can lead to low overhead for context switching; 2) the memory access within a CUDA block utilizes GPU caches. Both shared memory in the CUDA programming model (or local memory in OpenCL) and global memory with spatial/temporal locality within a CUDA block utilize caches. These memory accesses from a CUDA block are mapped into a CPU thread, so those memory accesses would also utilize CPU caches \cite{stratton2013performance}; 3) threads within a block have similar computation patterns, which makes them amenable to the SIMD instructions \cite{jeong2012performance,karrenberg2012improving} common in the current CPU architectures for further optimizations ~\cite{nuzman2006auto,nuzman2006multi,kar:hac11,nuzman2006autovectorization,maleki2011evaluation,porpodas2017supergraph,rosen2007loop}. The transformation is shown in Figure \ref{fig:SPMD}(b): for an original GPU kernel, the translator first splits it into different regions according to synchronization instructions and then wraps each region with a loop whose size equals to the GPU block size. However, this transformation was proposed based on early GPU programming models and cannot support several important features that were proposed in recent GPU programming models. One of the significant changes is the warp-level programming model in CUDA \footnote{Warp is now officially a part of the programming model instead of being a microarchitecture concept.}. And this new feature is critical for achieving high performance. hierarchical collapsing is proposed to support these warp-level features on CPUs. Although this might sound like a trivial extension, it is critical to identify new types of barriers in the warp level. And warp- and block-level barriers form a hierarchical relationship that complicates translating loops or branches into a CPU-friendly version. Throughout the paper, the focus is on the CUDA programming model. However, the same techniques would also be applicable to other SPMD programming models (e.g., OpenCL\cite{munshi2009opencl}, HIP~\cite{hip}, DPC~\cite{dpct}). Based on hierarchical collapsing, a framework, COX, is implemented that efficiently executes CUDA programs with the latest features on X86 devices. COX also uses SIMD instructions explicitly to take advantage of the hardware features in the latest CPUs.

 The main contributions of this paper as follows: 
\begin{itemize}
    \item propose hierarchical collapsing, which provides the correctness for mapping GPU programs that use warp-level functions to CPU programming models, and implement it with LLVM passes. 
    \item extend the Parallel Region concept into the Hierarchical Parallel Region to provide the correct translation when the GPU programs have warp-level functions.
    \item implement the COX framework, which executes CUDA source code on CPUs. The framework includes a new LLVM pass that contains hierarchical collapsing and a lightweight runtime system. \footnote{the frame will be released as an open source once the paper is accepted.}

\end{itemize}

\section{Background and Motivation\label{sec:background}}

\subsection{Running SPMD Programming Models on CPUs \label{sec:previous_concept}}

The basic mechanism to support SPMD programming models on CPUs is to map a thread block/work-group to a CPU thread and iterate for the thread block/work-group using a loop~\cite{kar:hac11, stratton2008mcuda,diamos2010ocelot,jaaskelainen2015pocl,karrenberg2012improving}.   Figure~\ref{fig:SPMD}(a) and (b) show the input and output of the basic process. This transformation has different names in different projects: microthreading~\cite{stratton2010efficient}, thread aggregation~\cite{zhang2013improving}, thread-fusion~\cite{diamos2010ocelot}, region-based serialization~\cite{stratton2013performance}, loop chunking~\cite{shirako2009chunking}, and kernel serialization~\cite{blomkvist2021cumulus}. In this paper, this transformation is called flat collapsing, as it uses a single loop to represent all threads within a block. The loop can be vectorized by the compiler, and multiple CPU threads (essentially multiple CUDA blocks) are executed in parallel on multiple cores using runtime system such as p-thread or openMP. When a GPU program contains synchronization primitives inside (e.g., {\tt synchthreads())}, the loop needs to be split (loop fission).

Below is some of the terminologies used in~\cite{jaaskelainen2015pocl, stratton2008mcuda, karrenberg2012improving} to support this transformation in general cases:
\begin{itemize}
    \item \textbf{Parallel Region (PR)}: These are the regions between barriers that must be executed by all the threads within a block before proceeding to the next region. Typically, loops are generated to wrap each PR. In Figure \ref{fig:SPMD}, {\tt statements-A} and {\tt statements-B} form two PRs. 
    \item \textbf{Extra barrier}: Unlike explicit barriers that are inserted by programmers (e.g., {\tt synchthreads()}), the flat collapsing inserts extra barriers that are necessary to define the Parallel Region. The flat collapsing groups instructions between barriers as PRs and wraps these PRs by loops. However, when barriers are present in the conditional statements, the situation becomes more complex. For example, to transform a CUDA kernel that has a barrier within an if–then construct, flat collapsing has to insert extra barriers in the original CFG so that it can get correct PRs in the further step.
\end{itemize}
The definition of Parallel Region in previous project for flat collapsing cannot support wrap-level features. The flat collapsing generates a single loop for each PR to simulate all threads within a block. This coarse-grain simulation cannot distinguish threads among different wraps. In this paper, an extension definition is proposed and used to support warp-level features. (See Section~\ref{sec:hierarchical_PR} for details.)
\subsection{Warp-level Programming Features \label{sec:cuda_feature}}
\subsubsection{Warp-Level Collectives\label{sec:cooperative_group_intro}}
CUDA provides a list of warp-level collective functions,~\footnote{\url{https://docs.nvidia.com/cuda/cuda-c-programming-guide/index.html##warp-shuffle-functions}} which are necessary when achieving high performance for reduction. This section introduces two of them that are commonly used in existing benchmarks.
\begin{itemize}
    \item \textbf{Warp shuffle}: In early versions, although most GPUs have local memory and global memory to support data exchange between threads, there was no efficient way to exchange data among threads within a warp. To efficiently exchange data that is stored in registers, CUDA provides a series of warp shuffle instructions. When the warp shuffle instructions are invoked, a thread can send its local data to another thread in the same warp. Warp shuffle can be used with warp vote to write more flexible programs.
    \item \textbf{Warp Vote}: Instead of only exchanging data among threads within a warp, warp vote instructions can directly make logical reductions (e.g., all, any) for the local variables, controlled by the {\tt mask} argument. These features, used with warp shuffle and cooperative group, are necessary to implement high-performance reduction kernels.
\end{itemize}
\subsubsection{Cooperative group\label{sec:cooperative_group_intro}}
In the early CUDA programming models, there are only two explicit groups of threads: block and grid. Users cannot easily organize a sub-group among a small group of threads within the block. NVIDIA proposed a new concept in CUDA 7.0 called {\tt cooperative group}. The corresponding instructions allow users to group threads within a block and this group can further be used for data exchange. There are two kinds of grouping strategies: static and dynamic. For the static grouping, it is known whether a thread belongs to a group in compile time (e.g., group threads with index 0 and 1), while for dynamic grouping, it can only be known during runtime (e.g., group all activated threads). 

\subsubsection{Limitation of COX}
This project focuses mainly on the compile-level transformation. Thus, only the static features can be addressed. The latest CUDA supports several dynamical features. For example, for warp-level collective operations, users can assign only a sub-group of threads within the wrap to do the collective operations. The sub-group is organized by a mask argument at runtime. For the cooperative group, users can also group all activated threads into a group at runtime. For these warp-level features, although these dynamic features provide more flexibility, they can be harmful for performance, as they may incur warp-divergence. Thus, most high-performance implementations~\cite{CUB,nai2015graphbig} use warp-level collectives and the cooperative group without warp-divergence. In the following sections, only the non warp-divergence use cases for these new features are of concern. For the same reason, only the aligned barriers\footnote{\url{https://docs.nvidia.com/cuda/parallel-thread-execution/index.html##parallel-synchronization-and-communication-instructions-bar}} are of concern. In other words, for a block/wrap barrier, it is assumed that all or none of the threads within a block/wrap can reach this barrier.
\subsection{Motivation}
\label{subsec:motiv} 

\begin{figure}[h]
    \centering
    \includegraphics[width=0.4\columnwidth,trim=4 12 4 4,clip]{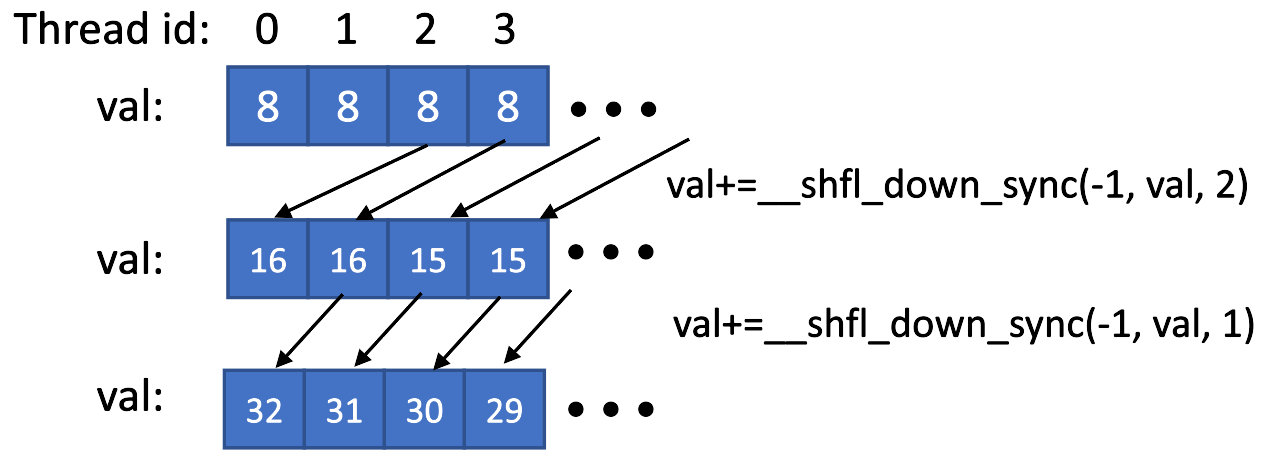}    
    \caption{Explanation of using {\tt shfl\_down\_sync} to implement reduction.}
    \label{fig:shfl_explanation}
\end{figure}

Section~\ref{sec:previous_concept} introduced the concepts of Parallel Region (PR) and extra barrier. This section discusses, with examples, the limitation of these concepts and how to extend them.
Assume the input kernel shown in Code~\ref{code:gpu_shfl},\footnote{This example is a simplified version of {\tt reduction\_kernel.cu} in CUDA 10.1 SDK.} and its block size is $b\_size$. The code accumulates the variable $val$ within the first warp and stores the accumulated value in the first thread of this warp. Figure~\ref{fig:shfl_explanation} illustrates an example of the last two iterations.

\begin{lstlisting}[caption={Input GPU reduction kernel},label={code:gpu_shfl},language=C]
  int val = 1;
  if (threadIdx.x < 32) {
    for (int offset = 16; offset > 0; offset /= 2)
        val += __shfl_down_sync(-1, val, offset);
  }
\end{lstlisting}

Although none of the existing projects can support this kernel, it is assumed that flat collapsing would generate code with the following steps:
\begin{itemize}
    \item group consecutive instructions between two barriers, and wrap each group with a for-loop whose length equals to the block size. In Code \ref{code:cpu_shfl_v1}, there are three groups that are separately wrapped by three for-loops in line 3, line 6, and line 9. Please note that the loop in line 5 is from the source input; 
    \item replace the use of {\tt threadIdx.x} by the loop iteration variable {\tt tx};
    \item replicate variables (e.g., {\tt val}) used by more than one group into an array form; 
\end{itemize}

Unfortunately, these steps are insufficient to generate the correct CPU program for this GPU kernel. The key reason is that there are implicit warp-level barriers derived from {\tt shfl\_down()}: each thread within a warp has to first calculate its {\tt val} and then invoke {\tt shfl\_down()} at the same time (see Section. \ref{subsec:warp} for details). These warp-level barriers are inside a branch of an if–then construct, which creates a  more complex situation. 
In previous projects, there are only block-level barriers, so it can safely be assumed that for each barrier instruction, all or no threads within the block can access it. Thus, it will just wrap the barrier instructions with for-loops with lengths {\tt b\_size}, and these for-loops are located in the branch of a if–then construct: if it is knwon at runtime that this barrier will be accessed, then the control flow will directly runs into this branch and execute the for-loop; or go to another branch otherwise. However, in the example, these warp-level barriers are only accessible for the threads in the first wrap. To get the correct result, one needs to not only wrap the barrier instructions with for-loops, but also replicate the control flow instruction in if–then construct (line 2 in Code~\ref{code:gpu_shfl}) and insert them into line 7 and line 10 in Code~\ref{code:cpu_shfl_v1}. With the above modifications,  Code~\ref{code:gpu_shfl} would become Code~\ref{code:cpu_shfl_v1}.
These transformations are quite complex, even for the demo example, not to mention implementing it in compilers used for all possible CFGs. Based on the above analysis, hierarchical collapsing is proposed which produces Code \ref{code:cpu_shfl_v3}. The concept is also illustrated in Figure~\ref{fig:SPMD}(c). 
Compared with Code~\ref{code:cpu_shfl_v1}, Code~\ref{code:cpu_shfl_v3} has two types of generated loops: the loop with induction variable {\tt wid} (line 4) is for the block level called {\em inter-warp loop} , while the inner loops with induction variable {\tt tx} (lines 5, 7, 12, 14) are for the warp level called {\em intra-warp loop}. With inter/intra warp loops, compared with a single-level loop for all threads within a block, the complexity of the generated code can be reduced: no longer needs to replicate and insert the if-then construct. Instead, it is maintained only with a simple loop peeling (line 10 in Code~\ref{code:cpu_shfl_v3}). With the low complexity, hierarchical collapsing can easily be implemented and integrated into compilers. In COX, hierarchical collapsing is implemented as a new LLVM pass to automatically transform the GPU kernels.
\begin{lstlisting}[caption={CPU warp shuffle program generated by flat collapsing},label={code:cpu_shfl_v1},language=C]
  int shfl_arr[32];
  int val[b_size];
  for (int tx = 0; tx < b_size; tx++)
    val[tx] = 1;
  for (int offset = 16; offset > 0; offset /= 2) {
    for (int tx = 0; tx < b_size; tx++)
      if (tx < 32)
        shfl_arr[tx] = val[tx];
    for (int tx = 0; tx < b_size; tx++)
      if (tx < 32) 
        if (tx + offset < 32)
          val[tx] += shfl_arr[tx + offset];
  }
\end{lstlisting}

\begin{lstlisting}[caption={CPU warp shuffle program by hierarchical collapsing},label={code:cpu_shfl_v3},language=C]
  int shfl_arr[32];
  int val[b_size];
  bool flag[32];
  for (int wid = 0; wid < b_size / 32; wid++) {
    for (int tx = 0; tx < 32; tx++)
      val[wid * 32 + tx] = 1;
    for (int tx = 0; tx < 32; tx++)
      flag[tx] = (wid * 32 + tx) < 32;
    // loop peeling
    if (flag[0]) {
      for (int offset = 16; offset > 0; offset /= 2) {
        for (int tx = 0; tx < 32; tx++)
          shfl_arr[tx] = val[wid * 32 + tx];
        for (int tx = 0; tx < 32; tx++)
          if (tx + offset < 32)
            val[wid * 32 + tx] += shfl_arr[tx + offset];
      }
    }
  }
\end{lstlisting}
Below are several details worth mentioning:
\begin{itemize}
    \item As the input CUDA kernel has a {\tt shfl\_down} inside an if-then construct, this is quite a complex situation: not all warps can access the implicit warp-level barriers derived from {\tt shfl\_down}. According to CUDA document, \footnote{\url{https://docs.nvidia.com/cuda/cuda-c-programming-guide/index.html##synchronization-functions}} for a given warp-barrier, none or all threads within a warp can access it. Thus, loop peeling (line 10 in Code \ref{code:cpu_shfl_v3}) is used to evaluate the condition of the if-then construct only for the first thread in the warp, and then all other threads in the warp just follow the same path. (See Section \ref{subsec:implicit} for details);
    \item Although only {\tt flag[0]} is needed, the instructions to calculate other elements in {\tt flag} are also executed. Because these instructions may have a side effect, to guarantee the correctness, they have to be executed even these outputs are not needed;
\end{itemize}


The rest of this paper is organized as follows: Section \ref{sec:IR_transformation} introduces the key part of the hierarchical collapsing. The runtime system is introduced in Section \ref{sec:runtime_system}. Section \ref{sec:experiment} shows the evaluation of COX with CUDA SDK, heter-Mark, and GraphBig benchmarks and the comparison of the performance with POCL and DPC, which are the state-of-the-art open source frameworks. Section \ref{sec:related_work} provides a survey that describes various attempts to migrate GPU programs to CPU devices. Finally, concluding thoughts are presented in Section \ref{sec:conclusion}.

\section{IR Transformation\label{sec:IR_transformation}}
\subsection{Overview of COX}
\begin{figure}[h]
    \centering
    \includegraphics[width=0.8\columnwidth,trim=4 12 4 4,clip]{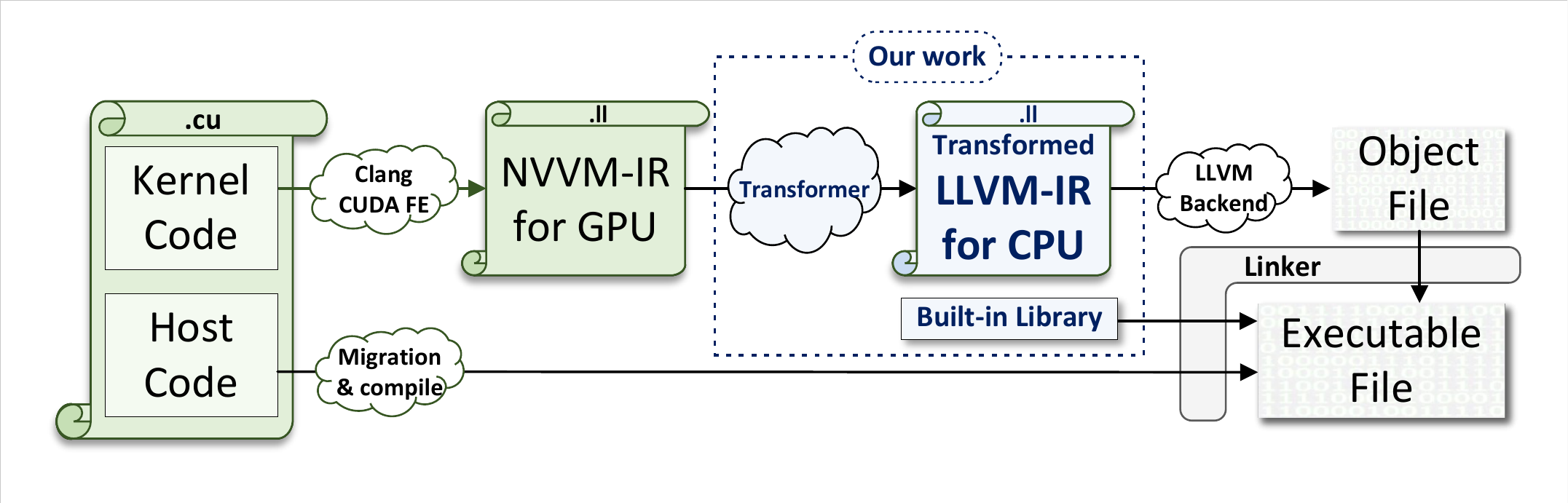}    
    \caption{COX pipeline for generating CPU executable files from CUDA kernel codes. }
    \label{fig:execute_pipeline}
\end{figure}

Figure~\ref{fig:execute_pipeline} shows an overview of the COX framework. At a high level, a CUDA kernel is compiled with Clang, which produces NVVM IR~\cite{NVVM} for the NVIDIA GPU. Then, COX transforms the NVVM IR into the CPU-friendly LLVM IR. The hierarchical collapsing~is implemented in this transformer. After that, the LLVM IR is linked with host programs and runtime libraries to generate a CPU-executable file. 

  \begin{lstlisting}[caption={CUDA Warp Vote example},label={code:cuda_warpvote},language=C]
__global__ void VoteAll(int *result) {
  int tx = threadIdx.x;
  result[tx] = __all_sync(-1, tx % 2);
}
\end{lstlisting}

  \begin{figure*}[ht]
        \centering
    	\begin{subfigure}[t]{0.3\textwidth}
			\includegraphics[width=\textwidth,trim=4 4 4 4,clip]{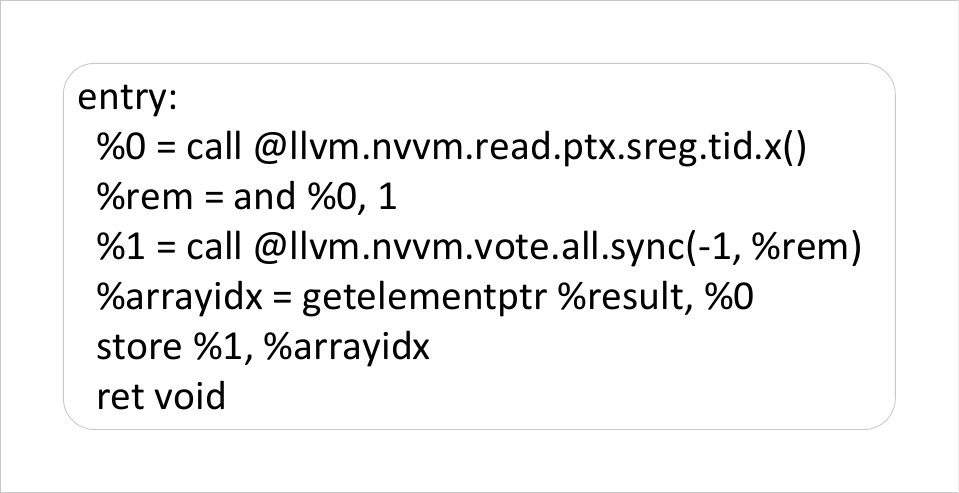}
			\caption{Original Code.}		
		\end{subfigure}
        \begin{subfigure}[t]{0.3\textwidth}
			\includegraphics[width=\textwidth,trim=4 4 4 4,clip]{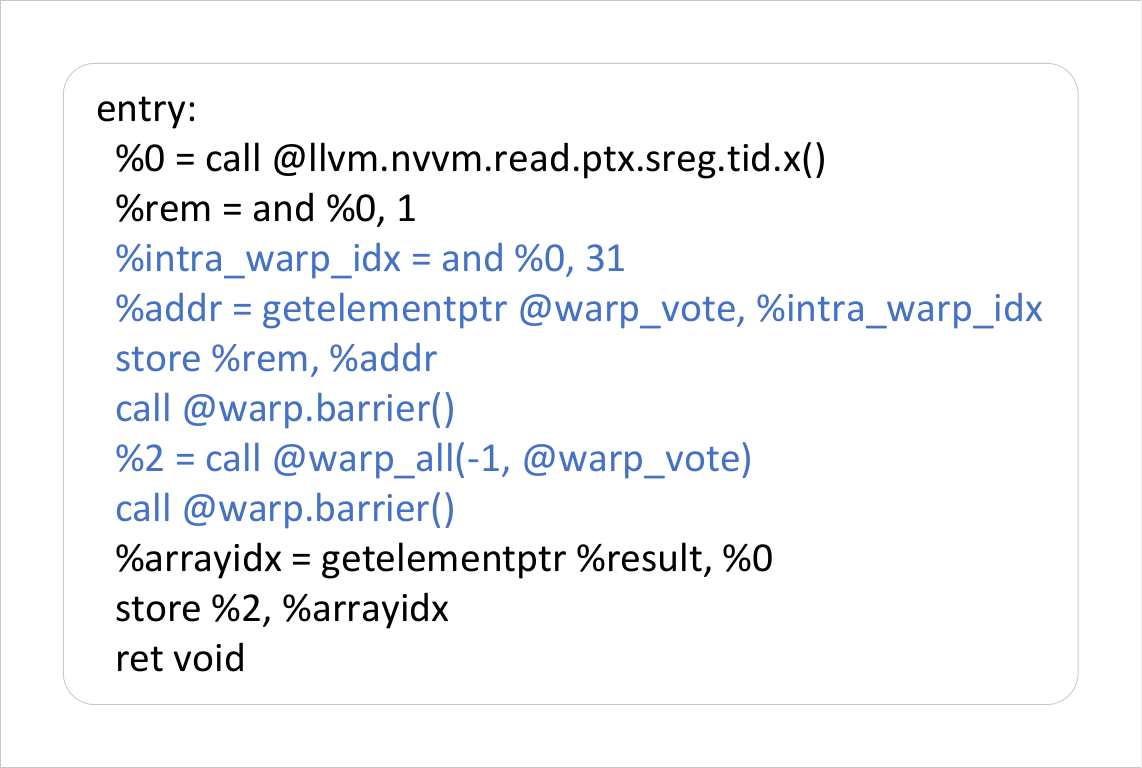}
			\caption{Step 1: Replace warp function.}			
		\end{subfigure}
        \begin{subfigure}[t]{0.3\textwidth}
			\includegraphics[width=\textwidth,trim=4 4 4 4,clip]{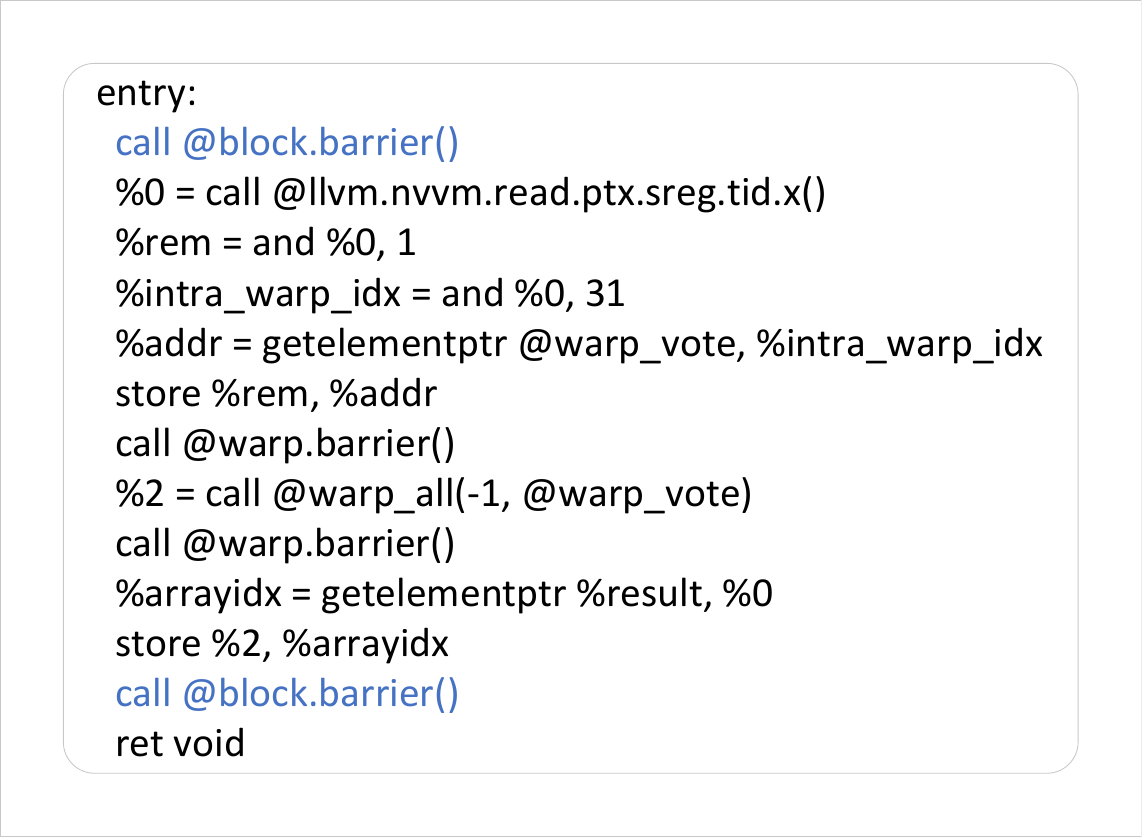}
			\caption{Step 2: Insert extra barriers.}			
		\end{subfigure}
    	\begin{subfigure}[t]{0.23\textwidth}
			\includegraphics[width=\textwidth,trim=4 4 4 4,clip]{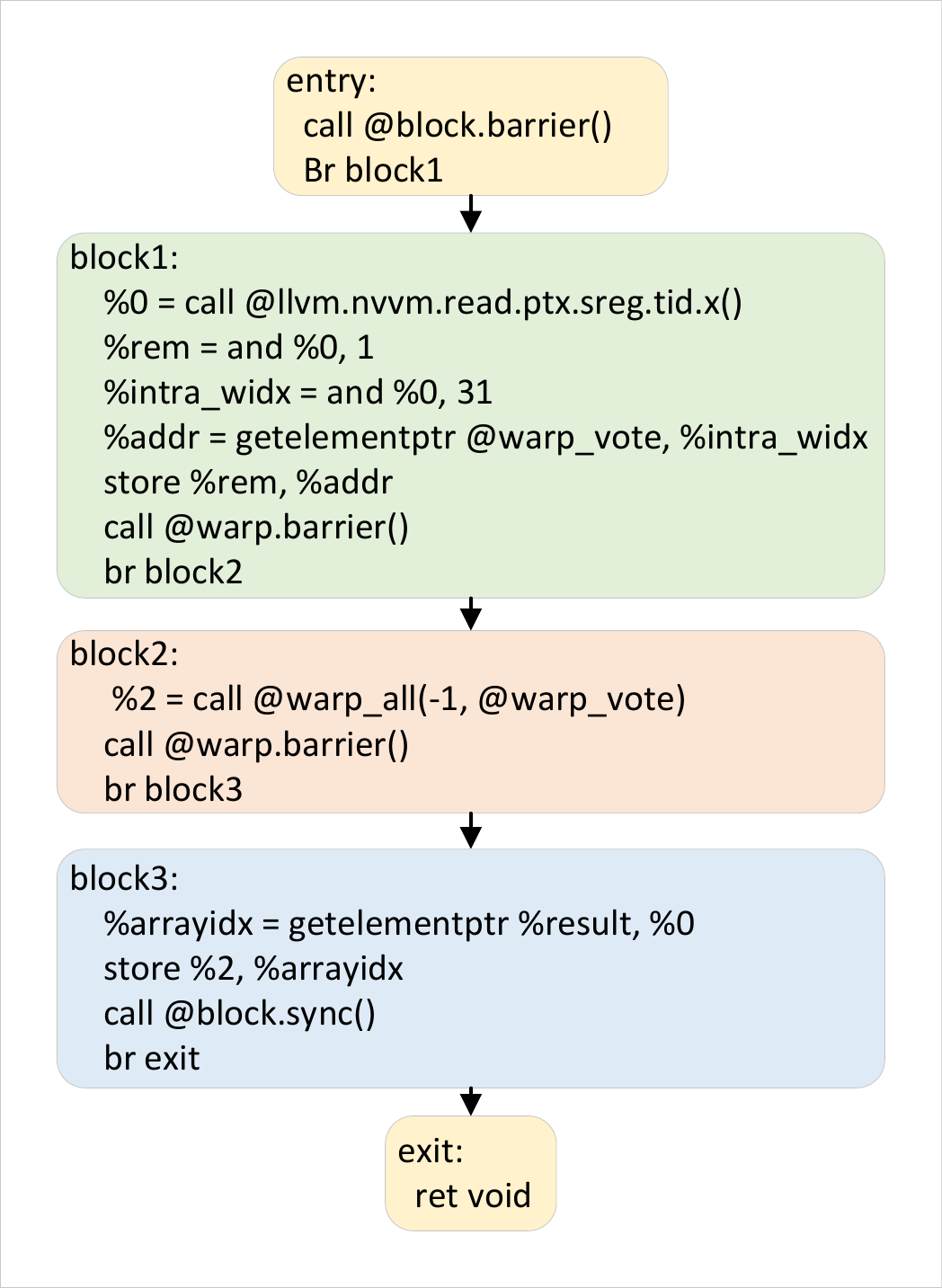}
			\caption{Step 3: Split blocks by barriers.}			
		\end{subfigure}
		\hspace{2mm}
        \begin{subfigure}[t]{0.49\textwidth}
			\includegraphics[width=\textwidth,trim=4 4 4 4,clip]{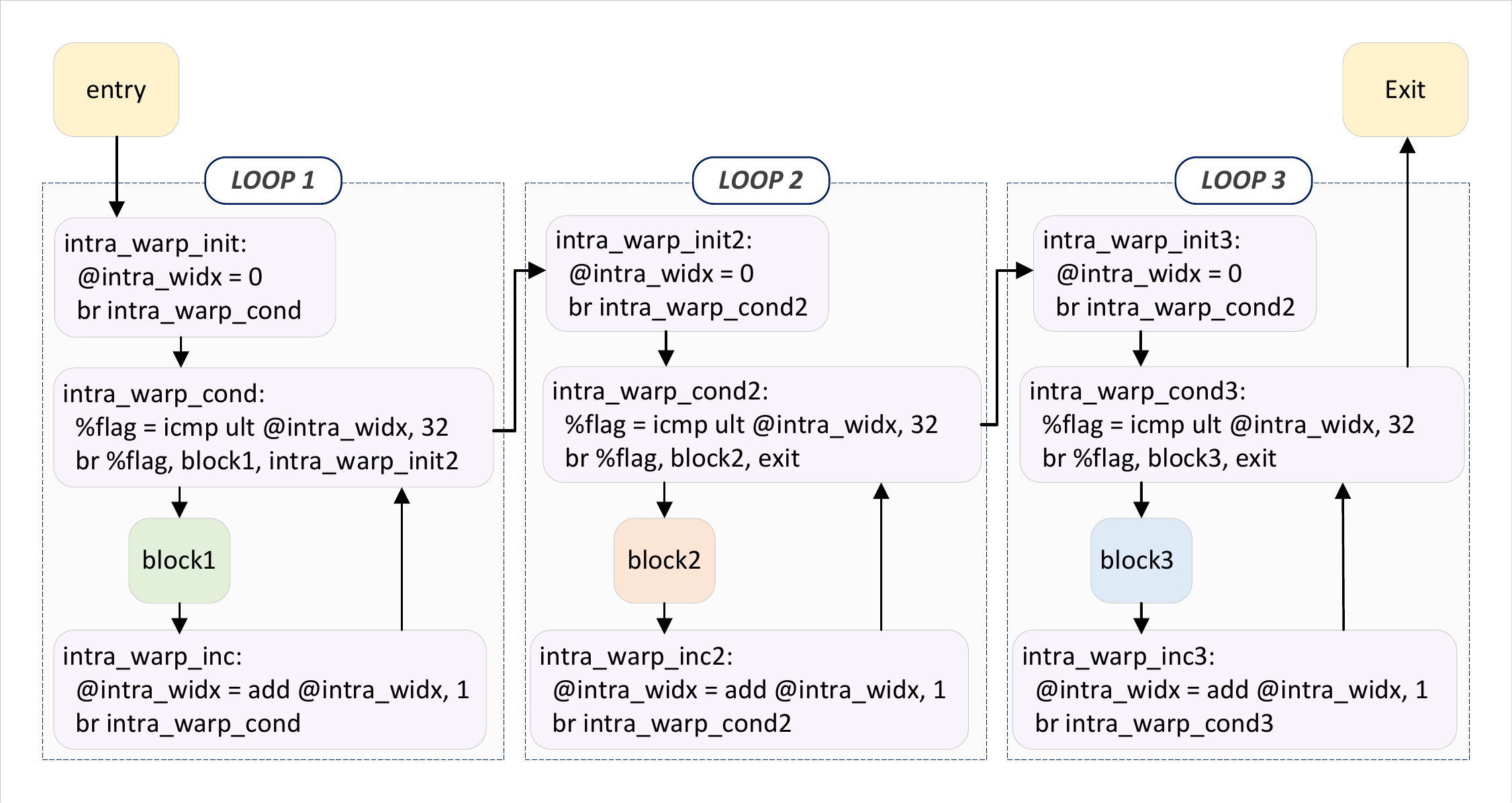}
			\caption{Step 4: Wrap the current CFG with intra-warp loop.}
		\end{subfigure}
		\hspace{2mm}
        \begin{subfigure}[t]{0.23\textwidth}
			\includegraphics[width=\textwidth,trim=4 4 4 4,clip]{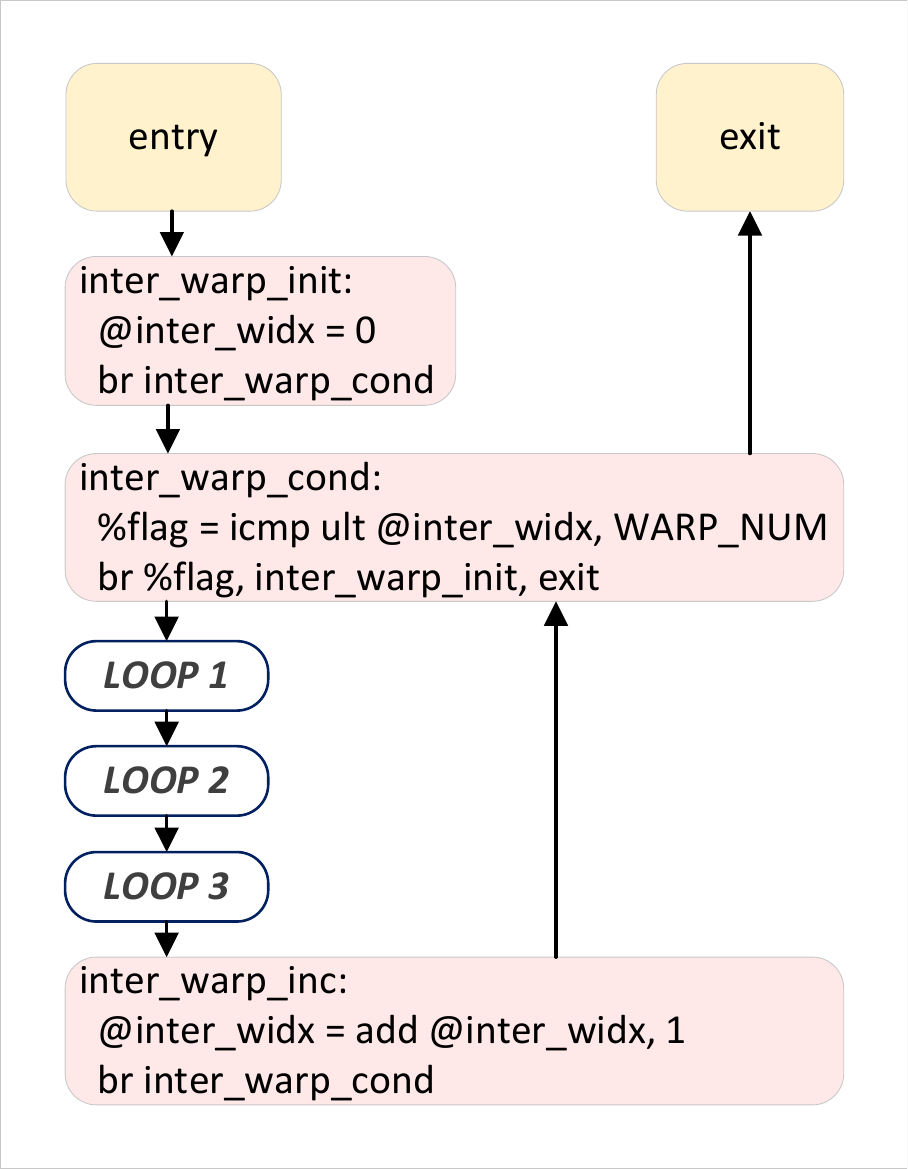}
			\caption{Step 5: Wrap the current CFG with inter-warp loop.}
		\end{subfigure}		
	\caption{Steps of NVVM to LLVM-IR transformer in Figure~\ref{fig:execute_pipeline}. }
	\label{fig:pipeline_demo}
	\end{figure*}

Figure~\ref{fig:pipeline_demo} shows an example of transforming CUDA kernels shown in Code~\ref{code:cuda_warpvote} into a LLVM-IR for CPU. First, warp-level functions are replaced with built-in functions defined in the runtime library, as shown in Step 1 (Section~\ref{subsec:warp}). Second, in Step 2, the extra barriers are identified and inserted (Section~\ref{subsec:implicit}). Last, through Steps 3 to 5, hierarchical parallel regions are identified, and intra/inter-warp loops are generated accordingly to create a CPU-friendly version (Section~\ref{subsec:split}). After Step 5, the generated LLVM IR will be compiled and linked with host programs and runtime libraries to generate a CPU-executable file.

\subsection{Support Warp-level Functions}
\label{subsec:warp} 
In a GPU architecture, when threads within a warp invoke the warp functions, the GPU will have internal communications to accumulate and/or communicate the local variables among the warp. To support these features on CPUs, the corresponding accumulation and communication need to be explicitly performed. This section describes how to support warp-level collectives. \\
In the initialization, COX allocates an array {\tt warp\_vote} with length 32. The {\tt warp\_vote} should be stored in CPU thread local memory, as a CPU thread is used to simulate a GPU block. Otherwise, if {\tt warp\_vote} is stored in global memory that is accessible to all CPU threads, a data race can occur when multi CPU threads read/write the {\tt warp\_vote} variable at the same time.
A GPU warp vote instruction is translated to the following CPU instructions: for threads within a warp, first, each thread stores its local flag into a different element in {\tt warp\_vote}. After all the elements are set, the result for this warp vote can easily be computed. The function {\tt warp\_all} is defined in a runtime library that will be linked at the final compilation. To utilize the computation resource of X86, {\tt warp\_all} is implemented with the AVX instructions. The benefits brought by AVX are evaluated in Section~\ref{sec:simd}. The ways to support warp shuffle are quite similar. See Code~\ref{code:gpu_shfl} and Code~\ref{code:cpu_shfl_v3} for example. \\
COX also needs to insert the implicit warp-level barriers when supporting these warp-level functions. As discussed in \cite{patel2021virtual}, two warp-level barriers are required: barriers for the Read-after-Write (RAW) hazard and barriers for the Write-after-Read (WAR) hazard.
The use of these two barriers is shown in Code~\ref{code:consecutive_warp}. There are two consecutive warp vote instructions and the inserted barriers 1) without the barriers for RAW hazard, a thread will invoke {\tt warp\_all} before other threads set {\tt warp\_vote[tx]} to 1 (the first vote) or 2 (the second vote); 2) without the barriers for the WAR hazard, a thread will set {\tt warp\_vote[tx]} to 2 before other threads invoke the first vote function. The use of consecutive warp-level collective functions is really common when implementing reduction.

\begin{lstlisting}[float,floatplacement=H,caption={Insert implicit Barriers to avoid RAW/WAR hazards},label={code:consecutive_warp},language=LLVM]
  ; the first warp vote instruction
  @warp_vote[tx] = 1;
  call @warp.sync() ; for RAW hazard
  %res1 = call @warp_all() ; read from @warp_vote 
  call @warp.sync() ; for WAR hazard
  ; the second warp vote instruction
  @warp_vote[tx] = 2;
  call @warp.sync() ; for RAW hazard
  %res2 = call @warp_all() ; read from @warp_vote 
  call @warp.sync() ; for WAR hazard
\end{lstlisting}

\subsection{Insert extra Barriers}
\label{subsec:implicit}
\begin{figure}[H]
        \centering
		\begin{subfigure}{.23\textwidth}
			\centering
			\includegraphics[width=\textwidth]{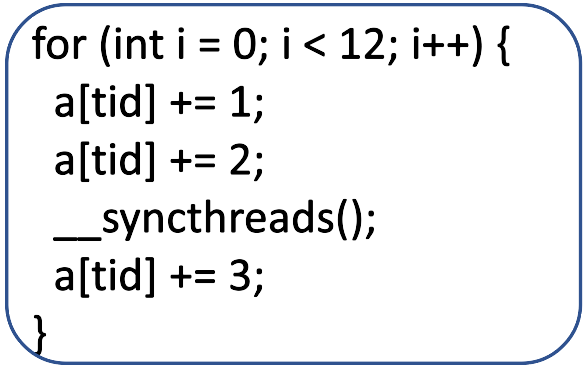}
			\caption{Input CUDA kernel}
		\end{subfigure}
		\begin{subfigure}{.22\textwidth}
			\centering
			\includegraphics[width=\textwidth]{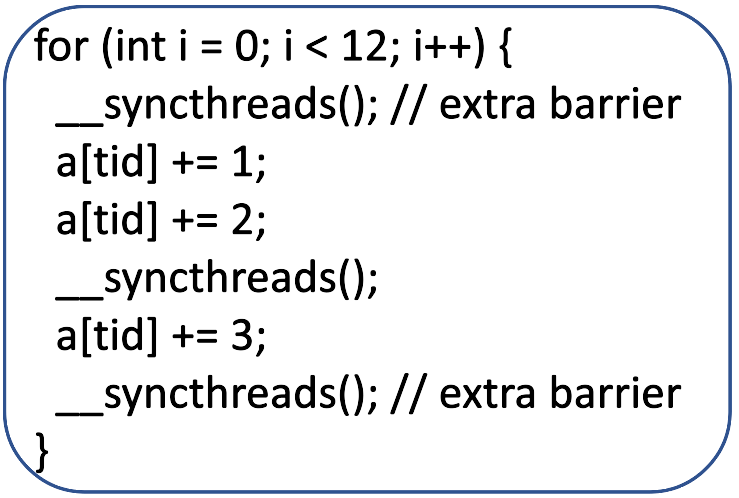}
		    \caption{After extra barrier insertion}
		\end{subfigure}
				\begin{subfigure}{.2\textwidth}
			\centering
			\includegraphics[width=\textwidth]{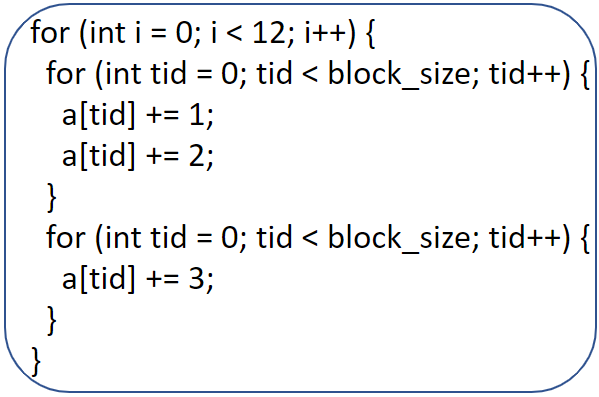}
		    \caption{After loop generation}
		\end{subfigure}
		\caption{An example of extra barriers needed for identifying PRs. a) the input CUDA kernel, which has a barrier in for-loop construct; b) as there is a barrier in the conditional statement, extra barriers are inserted to guide the generation of intra/inter-warp loops in future steps; c) according to the barriers, two PRs are identified and two for-loops are generated separately. Note, all transformations in COX are done in the LLVM IR level. This source code level example is only used for explanation.} 
	\label{fig:explicit_implicit_barrier}
	\end{figure}

In Steps 3, 4, and 5, hierarchical collapsing needs barrier information to identify the Parallel Region and generate intra/inter-warp loops accordingly. Thus, it is important to insert extra barriers that are not shown in the input GPU codes but necessary for identifying the Parallel Region. Some researchers \cite{karrenberg2012improving} proposes a similar concept and corresponding algorithm, but they cannot support warp-level functions. 
The extra barriers are sourced from barriers in conditional statements. An example is shown in Figure \ref{fig:explicit_implicit_barrier}.

To make the hierarchical collapsing work, extra block-level barriers are inserted at the beginning of the entry block and at the end of the exit block as POCL does\cite{jaaskelainen2015pocl}. \\
The two most common conditional statements are {\tt If-Then} construct and 2) {\tt For-loop} construct. 
\subsubsection{Barriers in if–then construct \label{sec:barrier_in_if}}
\begin{figure}[H]
    \centering
		\begin{subfigure}{.7\textwidth}
			\centering
			\includegraphics[width=\textwidth]{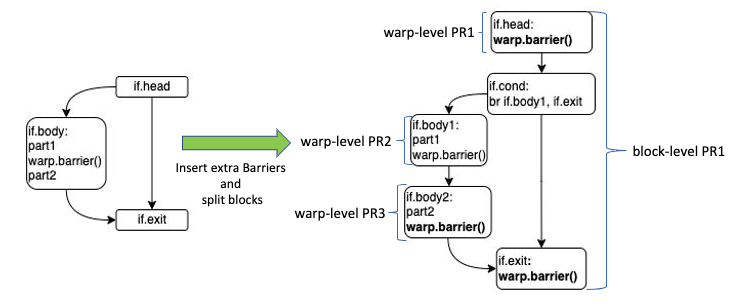}
			\caption{Barriers in if-then construct.}
		\end{subfigure}
		\\
		\begin{subfigure}{.7\textwidth}
			\centering
			\includegraphics[width=\textwidth]{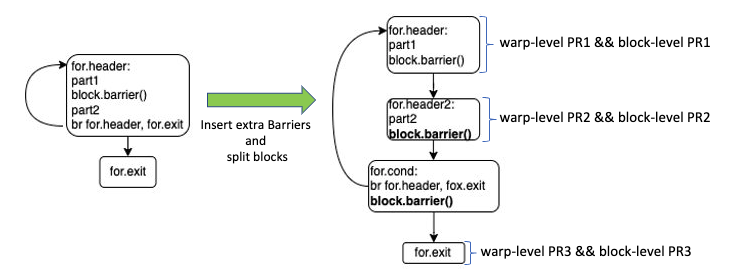}
		    \caption{Barriers in for-loop construct}
		\end{subfigure}
		\caption{After transformation, the inserted barriers are shown in bold. The PRs identified in further step are also shown in the figure.} 
	\label{fig:barrier_insert}
	\end{figure}

The CFG of a classical if–then construct is shown in left side of Figure \ref{fig:barrier_insert}(a). 
In the block {\tt if.body}, there is a barrier. According to \cite{CUDA-SYNC}, for a block/wrap barrier, none or all threads within the block/wrap can reach it\footnote{This rule does not exist for non-aligned barriers in CUDA, which is beyond the scope of this paper.}. Thus, COX can safely apply loop peeling on the CFG; COX peels the first thread to evaluate the branch direction and the rest of the threads within the warp/block can just follow the same direction. See Code \ref{code:cpu_shfl_v3} for a loop peeling example. The result after inserting extra barriers and block split is shown in the right side of Figure~\ref{fig:barrier_insert}(a). Several details are worth mentioning:
\begin{itemize}
    \item insert extra barriers with the same type as the barrier in {\tt if.body}. In the example, there is a warp barrier in {\tt if.body}, thus, hierarchical collapsing also inserts warp barriers as extra barriers;
    \item after transformation, all blocks will be wrapped by intra-warp loops, except {\tt if.cond}, which is used for loop peeling;
    \item {\tt if.cond} should contain only a single conditional-branch instruction and it should not have any side-effect. In Figure \ref{fig:barrier_insert}(a), all computation instructions are put into {\tt if.head} so that they are executed {\tt b\_size} times, as the original GPU program does.
\end{itemize}
    
The detailed algorithm for inserting extra barriers derived from barriers in if-then construct is described in Algorithm \ref{alg:insert_if_barrier}. COX has to do some additional checking to avoid an infinite loop caused by a for-loop construct in CFG. For simplicity this checking part is not shown in Algorithm \ref{alg:insert_if_barrier}.


  \begin{algorithm}[htbp]
  \caption{Transformation for inserting extra barriers for barrier in if-then construct}
  \label{alg:insert_if_barrier}
  \begin{algorithmic}[1]
    \Require $K$: The CFG for the input kernel
    \Require $PDT$: The Post Dominator Tree for the input CFG
    \Require $DT$: The Dominator Tree for the input CFG

    \State $conditional\_block \gets  []$
    \LeftComment {Find all barriers in if-body construct}
    \ForAll{$block \in K$}
        \If{$has\_barrier(block)$}
            \If{$!PDT.dominates(block, K.entry)$}
                \State $conditional\_block.insert(block)$
            \EndIf
        \EndIf
    \EndFor
        
    \LeftComment {Insert extra barriers}
    \ForAll{$block \in conditional\_block$}
        \State $NearestEntry \gets block.precessor$
        \While{$PDT.dominates(block, NearestEntry)$}
            \State $NearestEntry \gets NearestEntry.precessor$
        \EndWhile
        \LeftComment {Insert barrier in the end of if-head}
        \State $insert\_barrier\_before(precessor.terminator)$
        \State $pre\_successor \gets block$
        \State $successor \gets block.successor$
        \While{$DT.dominates(block, successor)$}
            \State $pre\_successor \gets successor$
            \State $successor \gets successor.successor$
        \EndWhile
        \LeftComment {Insert barrier in the beginning of if-exit}
        \State $insert\_barrier\_before(successor.begin)$
        \LeftComment {Insert barrier in the end of if-body}
        \State $insert\_barrier\_before(pre\_successor.terminator)$
        
        \LeftComment {Inserted extra barriers may generate another if-then construct that contains barriers}
        \If{$!PDT.dominates(NearestEntry, K.entry)$}
         \State $conditional\_block.insert(NearestEntry)$
        \EndIf
    \EndFor
    
  \end{algorithmic}
  \end{algorithm}

\subsubsection{Barriers in for-loop construct}
\label{subsubsec:barrier_for}
Although CUDA supports several loop styles (e.g., for-loop, while-loop, do-while-loop), after LLVM's transformation, all loops will be simplified to the canonical format, which 1) has single latch; 2) has the loop headers dominating all exit blocks.\footnote{A latch is a node in the loop that has an edge to the loop header. An exiting edge is an edge from inside the loop to a node outside of the loop. The source of such an edge is called an exiting block, its target is an exit block.\cite{llvm-loop}} Thus, COX only needs to concern these canonical loops.

COX inserts extra barriers before/after the branch instructions (back edge of the loop). Figure~\ref{fig:barrier_insert}(b) shows the example of inserting extra barriers for a for-loop construct which contains a block barrier. The same as with an if–then construct, all these inserted extra barriers (shown by bold text in the figure) should have the same type as with the barriers in {\tt for.header} (block barrier in the example).

\subsubsection{Supporting other conditional statements} 
CUDA is a high-level flexible language; even a single concept can generate quite a different CFG. For example, the loop concept can be implemented by different CFGs, such as do-while-loop, while-loop, and for-loop. However, with the existing LLVM transformations, COX can automatically convert the input CFGs to canonical formats and only focus on these canonical formats in the above discussion. Below are some important features in the canonical format:
\begin{itemize}
    \item Each branch instruction has only two successors; most input CFGs already have this feature, except the CFG which uses switch-case construct. For these exceptions, COX uses LLVM's $lowerswitch$ transformation to convert the switch-case construct into if–then constructs.
    \item All loops are in canonical format that 1) they all have pre-headers; 2) each loop has only a single latch, in other words, a single backedge; and 3) the loop header will dominate all exit blocks. COX calls LLVM's $loop-simplify$ transformation to translate the input CFG loops to the canonical format. 
\end{itemize}

\subsection{Split Blocks Before/After Each Barrier}
\label{subsec:split} 

As the instructions before/after a barrier in a block need to be wrapped by different intra/inter-warp loops, in this step, COX splits the blocks that have barriers inside. See Step 3 in Figure \ref{fig:pipeline_demo} for an example. 

\subsection{Hierarchical Parallel Region \label{sec:hierarchical_PR}}

\begin{figure}[htbp]
    \centering
    \includegraphics[width=90mm]{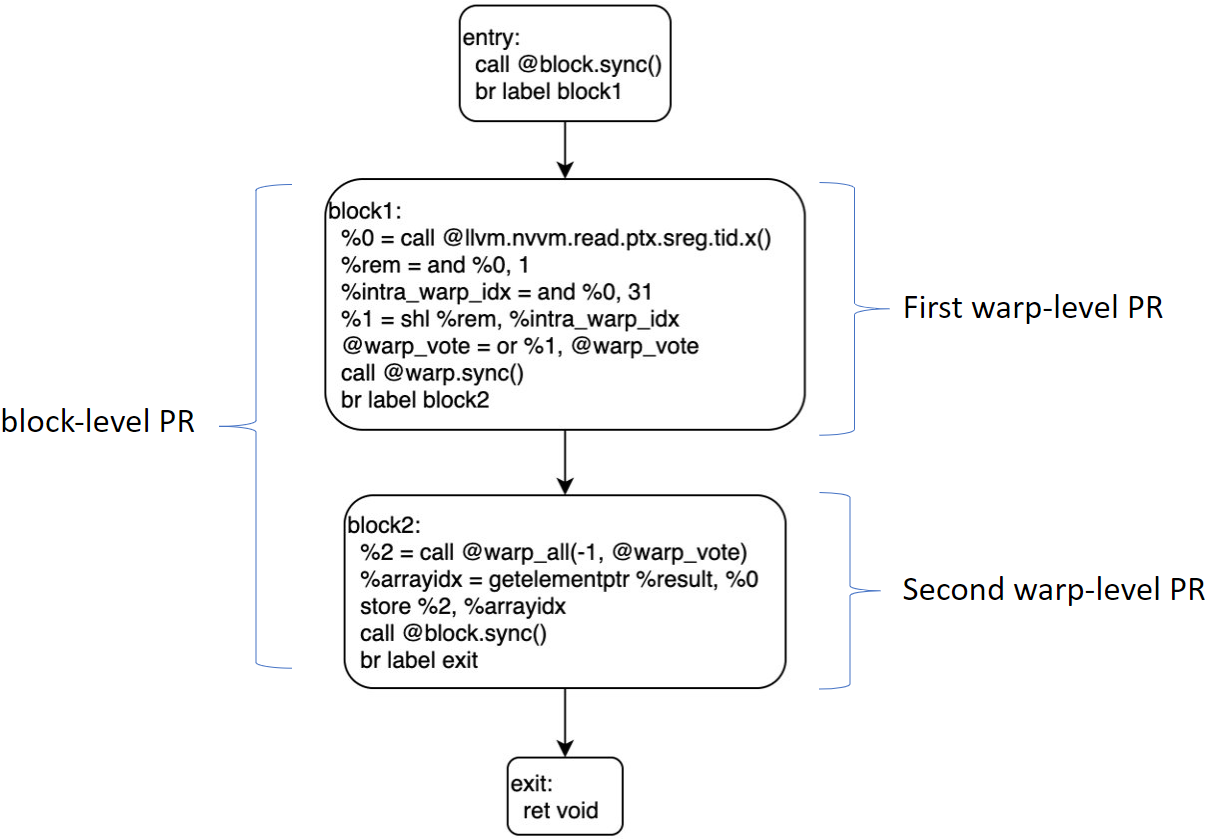}
    \caption{As there is a warp barrier in {\tt block1}, after transformation, there are two warp-level PR (\{ block1\},\{block2\}) and a single block-level PR (\{block1, block2\}).}
    \label{fig:PR_example}
    \end{figure}

As discussed in Section~\ref{subsec:motiv}, each parallel region (PR) becomes a for-loop. Due to the warp-level functions, COX has to generate two kinds of for-loops: inter-warp loop and intra-warp loop. Thus, COX needs two kinds of Parallel Regions: 1) warp-level Parallel Region, which will be wrapped by intra-warp loop and 2) block-level Parallel Region, which will be wrapped by inter-warp loop. It is obvious that a warp-level PR will always be a subset for a block-level PR (a GPU warp is always within a GPU block); thus, the new concept is called a Hierarchical Parallel Region. An example of a Hierarchical Parallel Region is shown in Figure \ref{fig:PR_example}.

Thus, the rest of the steps are for finding the block/warp level PRs and wrapping them with inter/intra-warp loops.
The algorithm (Alg. \ref{alg:intra_warp_PR}) is used for finding the set of warp-level PRs. The algorithm for finding the block-level PRs is very similar, except it only concerns the block barrier.
COX cannot find the PR for the warp level and block level simultaneously: COX first finds all warp-level PRs and generates intra-warp loops to wrap these PRs. Then, COX finds the block-level PRs in the new CFG and wrap them with inter-warp loops.
    
  \begin{algorithm}[htbp]
  \caption{Find all warp-level PRs}
  \label{alg:intra_warp_PR}
  \begin{algorithmic}[1]
    \Require $K$: The CFG after Step3
    \Ensure $PR\_set$: The set of PRs.
    
    \State $PR\_set \gets \{\}$
    \State $end\_block \gets  []$
    \ForAll{$block \in K$}
        \If{$block$ contains warp/block barrier}
            \State $end\_block.insert(block)$
        \EndIf
    \EndFor
    
    \LeftComment {Find the PR that $block$ belongs to}
    \ForAll{$block \in end\_block$}
        \If{block has more than one precessors}
        \State $continue$ \Comment{This is the exit of an if-then construct}
        \EndIf
        \State $PR \gets \{block\}$
        \State $pending\_block \gets block.precessors$
        \While{$!pending\_block.empty()$}
            \State $current \gets pending\_block.front()$
            \State $pending\_block.pop()$
            \If{has visited $current$}
                \State $continue$
            \EndIf
            \If{$current$ has warp/block barriers}
                \State $continue$
            \EndIf
            \State $PR.insert(current)$
            \State $pending\_block.insert(current.preprocessors)$
        \EndWhile
        \LeftComment {Blocks for loop peeling do not belong to any PR}
        \If{PR only has a single block which only contains a conditional branch}
        \State $continue$
        \EndIf
        \State $PR\_set.insert(PR)$
    \EndFor
  \end{algorithmic}
  \end{algorithm}

\ignore{
In the following paragraphs, we prove that the Alg. \ref{alg:intra_warp_PR} can return a valid set of PRs for group {\tt G} that satisfy:
\begin{itemize}
    \item \textbf{There is one and only one block in the set that has a barrier whose scope contains {\tt G}}: We first insert a block BB that has a barrier whose scope contains {\tt G} (line12 in Alg. \ref{alg:intra_warp_PR}). Then we continue visit BB's precessors and only insert these blocks that do not have barrier whose scope contains {\tt G} into the set.
    \item \textbf{The block that has a barrier whose scope contains {\tt G} post-dominates all other blocks in the set}: Following the above definition, BB is the first block inserted into the PR, and all other blocks are BB's ancestors. If there is a block PB that can not be post dominated by BB, there must be a for-loop; BB is within the for-body and PB is inserted into the PR after we visit the backedge of that for-loop. However, we have already inserted extra barrier at the beginning of each for-body, so we stop before reaching the backedge.
    \item \textbf{We cannot insert any other block not contained in the PR and get a new PR}: If there is a block {\tt BB} not contained in the current PR that can be inserted into the current PR and still a valid new PR, we know {\tt BB} is post-dominated by {\tt tail}. \textcolor{red}{(Maybe we should just remove these proof due to page limitation)}
\end{itemize}
}

\subsection{Wrap PR with For-Loop \label{subsec:warp_with_for}}

In this step, COX wraps warp/block-level PRs by intra/inter-warp loop. Please see Figure \ref{fig:pipeline_demo}(e)(f) for an example. 
Although this step is quite straightforward, it requires proving the correctness~\footnote{Due to page limitation, we move the proof into Appendix}: after inserting intra/inter-warp loops, each instruction from the input GPU kernel is executed {\tt b\_size} times ({\tt b\_size} is the block size), except the instructions used for loop peeling. 

Finally, after adding intra/inter-warp loops, some local variables are needed to be replicated: for local variables that are used in several warp-level PRs but only used in a single block-level PR, they are replicated with an array of length 32. For local variables that are used among different block-level PRs, they are replicated by an array of length equals to block size.

\section{Runtime System\label{sec:runtime_system}}
\begin{figure}[htbp]
    \centering
		\begin{subfigure}{.5\textwidth}
			\centering
			\includegraphics[width=\textwidth]{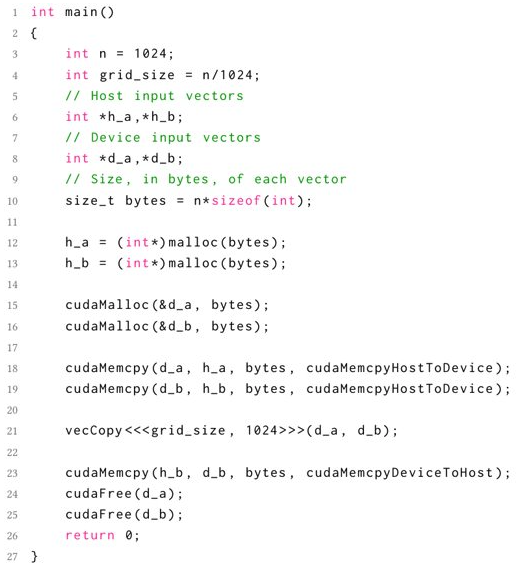}
			\caption{CUDA host code.}
		\end{subfigure}
		\\
		\begin{subfigure}{.9\textwidth}
			\centering
			\includegraphics[width=\textwidth]{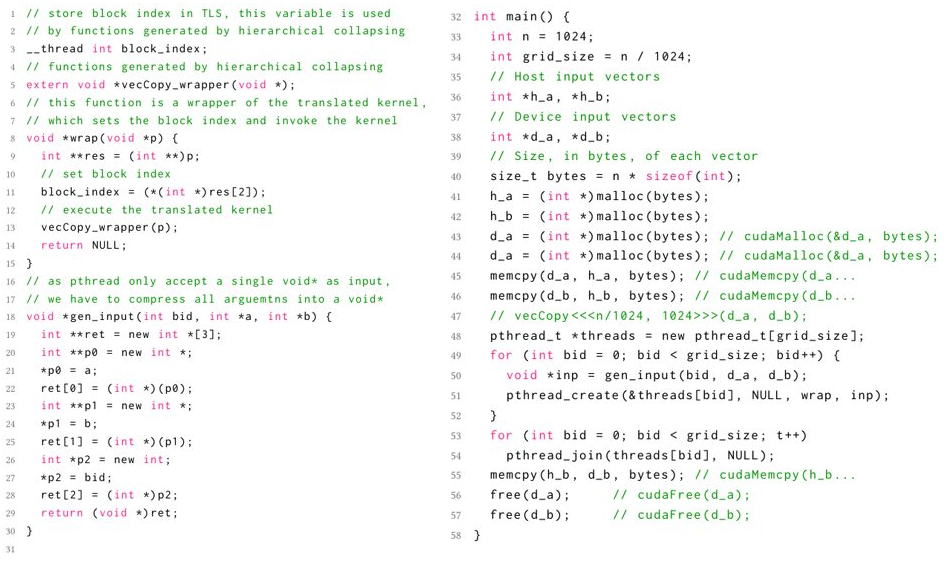}
		    \caption{Migrated COX host code.}
		\end{subfigure}
		\caption{The host code in COX is similar with the original CUDA host code. The automatic migration from CUDA host code to COX host code is open for future work.} 
	\label{fig:vecCopy_example}
	\end{figure}

The above section describes only the CUDA device part. As for the CUDA host part which involves memory allocation, memory transfer, and kernel launch, these features has to be manually migrated. The automatic translation from CUDA host code to CPU is left for future work. In the runtime system, p-thread is used for multi-threads. In this paper, both host and device are x86; thus, CUDA malloc and memcpy are replaced by C malloc and memcpy. In Figure \ref{fig:vecCopy_example}(a), a CUDA host example is presented for vector copy. The migrated COX host code is shown in Figure \ref{fig:vecCopy_example}(b), with the corresponding CUDA operations recorded in the comments. Compared with the CUDA host program, the COX host program has the following differences: 1) COX uses thread-local variable {\tt block\_index} to store the block index, and the block index is explicitly set during invocation; 2) COX replaces the CUDA memory operations with corresponding CPU operations; 3) COX uses pthread fork/join to replace kernel launch in CUDA. There are several potential optimization for the runtime system, such as using thread-pool instead of fork/join for kernel launching and using regular expression or LLVM transformation to automatically generate COX host programs from the CUDA source code. These optimizations are beyond the scope of this paper and are open for future research. 

The following steps make up the workflow for COX: It 1) compiles the input CUDA source code with Clang and gets the NVVM IR of kernel functions and; 2) transforms the NVVM IR with hierarchical collapsing; 3) links the transformed kernel with the COX host program (manually migrated from CUDA host program) and generates the CPU-executable file. COX has two modes: 1) normal mode to maintain the runtime configuration as variables (e.g., grid size, block size) and 2) JIT mode to compile the program with the given runtime configuration. In the normal mode, COX only needs to compile programs once, and it can be used for different runtime configurations. In JIT mode, a program has to be recompiled when executed with different configurations. Although the JIT mode requires recompiling, in some cases, it can generate higher-performance programs, as it provides more opportunities for optimizations. For more details, please see Section~\ref{sec:jit_mode}.

\section{Experimental Results\label{sec:experiment}}
To verify the correctness and performance, this section describes the experiments for supporting the CUDA kernel in several benchmarks on X86 and ARM architectures. Although several frameworks also support executing CUDA on CPU, most of them were developed decades ago and cannot really support programs that use new CUDA features. Below are the hardware and software environments for the experiments:
\begin{itemize}
    \item Software: Ubuntu 18.04, LLVM 10.0, gcc 7.5.0, CUDA 10.1, POCL 1.4, DPCT~\cite{dpct} Ver. : 2021.3.0.
    \item X86-Hardware: 8 x Intel(R) Xeon(R) Silver 4210 CPU @ 2.20GHz
     
    \item ARM-hardware: 48 x ARM(R) A64FX CPU @ BogoMIPS 200.00
    \item Benchmarks: CUDA SDK 10.1, Hetero-mark~\cite{sun2016hetero}, GraphBig~\cite{nai2015graphbig};
    \item Time: average time for running more than 1000 times; fork/join time of threads is also included.
\end{itemize}

\subsection{Coverage}
Table~\ref{table:cuda_sdk} analyzes examples in CUDA SDK10.1 that use but do not require special hardware support  (e.g., tensorcore, unified memory). 
POCL and DPCT are chosen for the coverage comparisons since they are the currently activated projects that support executing CUDA programs on CPUs. As POCL is designed for OpenCL and cannot directly execute CUDA programs, a third-party translator~\cite{han2021supporting} is used to support executing CUDA with POCL. Besides, although POCL has both compilation and runtime parts, only the compilation is used in this experiment. For POCL evaluation, the GPU programs is compiled by POCL and then executed on COX. Thus, it's fair to compare the execution time to show the results of compilation and avoid the effect of runtime system. \\
As shown in the table, the existed frameworks can only automatically support at most 21 kernels (coverage=68\%). Those failed kernels are using new CUDA features. On the other hand,  COX supports 28 kernels (coverage=90\%). 

\begin{table}[htp]
\tiny
\begin{tabular}{L{0.27\columnwidth}L{0.27\columnwidth}C{0.05\columnwidth}C{0.05\columnwidth}C{0.05\columnwidth}C{0.05\columnwidth}}
\hline
kernel name               & features                            & POCL              & DPCT          & COX         \\ \hline
initVectors               &                                     & \greenv           & \greenv       & \greenv                      \\ 
gpuDotProduct             & warp cooperative group                      & \redx             & \redx         & \greenv                      \\
gpuSpMV                   &                                     & \greenv           & \greenv       & \greenv                      \\ 
r1\_div\_x                &                                     & \greenv           & \greenv       & \greenv                      \\ 
a\_minus                  &                                     & \greenv           & \greenv       & \greenv                      \\ 
gpuConjugateGradient      & grid sync          & \redx             & \redx         & \redx                       \\ 
multigpuConjugateGradient & multi grid sync                     & \redx             & \redx         & \redx                       \\ 
MatrixMulCUDA             &                                     & \greenv           & \greenv       & \greenv                      \\ 
matrixMul                 &                                     & \greenv           & \greenv       & \greenv                      \\ 
copyp2p                   &                                     & \greenv           & \greenv       & \greenv                      \\ 
reduce0                   &  block cooperative group            & \redx             & \greenv       & \greenv               \\ 
reduce1                   &  block cooperative group            & \redx             & \greenv       & \greenv                  \\ 
reduce2                   &  block cooperative group            & \redx             & \greenv       & \greenv                  \\ 
reduce3                   &  block cooperative group            & \redx             & \greenv       & \greenv                  \\ 
reduce4                   & warp cooperative group        & \redx             & \redx         & \greenv                  \\ 
reduce5                   & warp cooperative group          & \redx             & \redx         & \greenv                  \\ 
reduce6                   & warp cooperative group          & \redx             & \redx         & \greenv                  \\ 
shfl\_intimage\_rows      & warp shuffle                        & \redx             & \greenv       & \greenv                  \\ 
shfl\_vertical\_shfl      & warp shuffle                        & \redx             & \greenv       & \greenv                  \\ 
shfl\_scan\_test          & warp shuffle                        & \redx             & \redx*       & \greenv                  \\ 
uniform\_add              &                                     & \greenv           & \greenv       & \greenv                  \\ 
reduce                    & warp cooperative group                      & \redx             & \redx         & \greenv                  \\ 
reduceFinal               & warp cooperative group                      & \redx             & \redx         & \greenv                  \\ 
simpleKernel              &                                     & \greenv           & \greenv       & \greenv                  \\ 
VoteAnyKernel1            & warp vote                           & \redx             & \greenv       & \greenv                  \\ 
VoteAllKernel2            & warp vote                           & \redx             & \greenv       & \greenv                  \\ 
VoteAnyKernel3            & warp vote                           & \redx             & \greenv       & \greenv                  \\ 
spinWhileLessThanone      &                                     & \greenv           & \greenv       & \greenv                  \\ 
matrixMultiplyKernel      &                                     & \greenv           & \greenv       & \greenv      \\ 
vectorAdd                 &                                     & \greenv           & \greenv       & \greenv     \\ 
filter\_arr               & activated thread sync               & \redx             & \redx         & \redx        \\ \hline
Coverage                  &                                     & 39\%              & 68\%          & 90\%       \\ \hline

\end{tabular}
\caption{Coverage of COX compared to other frameworks. *enabled by manual code migration\cite{tsai2021porting}}
\label{table:cuda_sdk}
\end{table}

\begin{figure}[htbp]
    \centering
    \includegraphics[width=75mm]{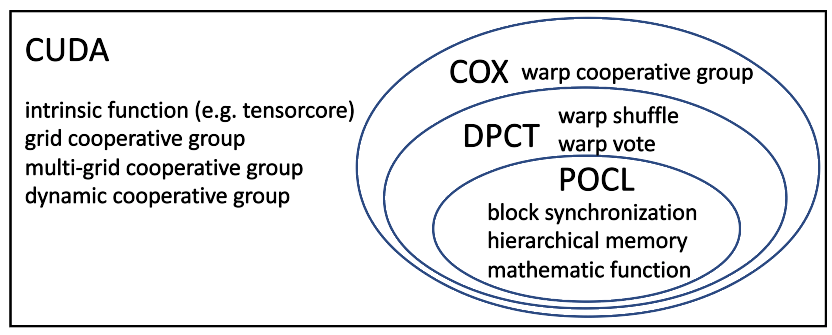}
    \caption{The CUDA features supported by POCL, DPCT and COX.}
    \label{fig:venn_diagram}
\end{figure}

The CUDA features supported by POCL, DPCT and COX is also shown in Figure \ref{fig:venn_diagram}.

Although the coverage with COX can be significantly improved, there are still three kernels that cannot be supported yet. $gpuConjugateGradient$ and $multiGpuConjugateGradient$ rely on synchronization between different grids and devices, which utilize the grid cooperative group and multi-grid cooperative group separately. $filter\_arr$ uses a dynamic cooperative group: it dynamically groups all activated threads. As discussed in Section~\ref{sec:cooperative_group_intro},  all these features should be supported at the runtime level: frameworks should schedule threads accordingly at runtime, and each thread can only know whether it is activated during runtime. Supporting runtime features is included for future work.

\subsection{Performance \label{sec:perf}}

\begin{figure}[htbp]
    \centering
    \includegraphics[width=0.8\columnwidth]{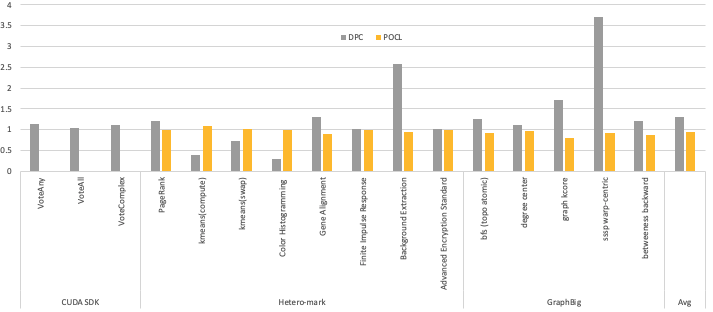}
    \caption{The normalized execution time of POCL and DPC on X86 architecture. The normalized execution time is the POCL/DPC execution time divide by COX's execution time. Thus, the COX's normalized execution time is always 1.}
    \label{fig:perf}
\end{figure}

Figure~\ref{fig:perf} shows a performance comparison of POCL, DPC, and COX on CUDA SDK, Hetero-Mark, and GraphBig benchmark on X86 architecture. 

In most cases, COX and POCL have close execution time. Thus, the POCL's normalized execution times are always close to 1. However, DPC's normalized execution time has large variance. This is due to: 1) DPC's most optimizations are for multiple block cases, which are runtime optimization. While in the evaluation, to shown the compile-level optimization, there is only a single block in each application; 2) DPC has optimizations on new Intel CPUs, while POCL and COX do not have special optimizations for the new Intel architectures.

\begin{figure}
    \centering
    \includegraphics[width=0.8\columnwidth]{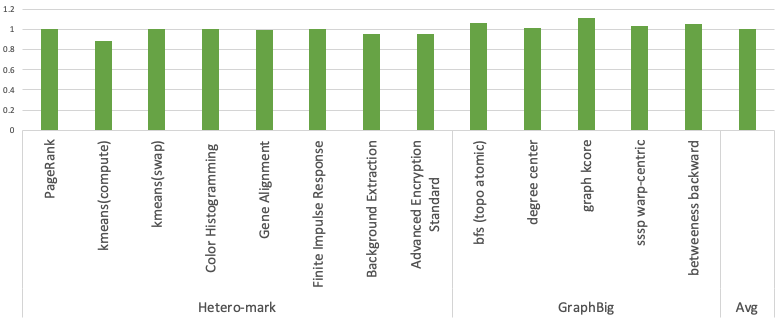}
    \caption{The normalized execution time of COX (normalized with POCL execution time) on ARM CPU. In all cases, COX and POCL has close performance. DPC does not support ARM CPU.}
    \label{fig:perf_arm}
\end{figure}

The evaluation results for ARM CPU with AArch64 architecture is shown in Figure \ref{fig:perf_arm}. As DPC does not support ARM CPU, only POCL and COX are evaluated. The performance between COX and POCL are close among all experiments. 

\subsubsection{Performance effects of flat collapsing vs. hierarchical collapsing}

\begin{figure}[htbp]
    \centering
    \includegraphics[width=0.7\columnwidth]{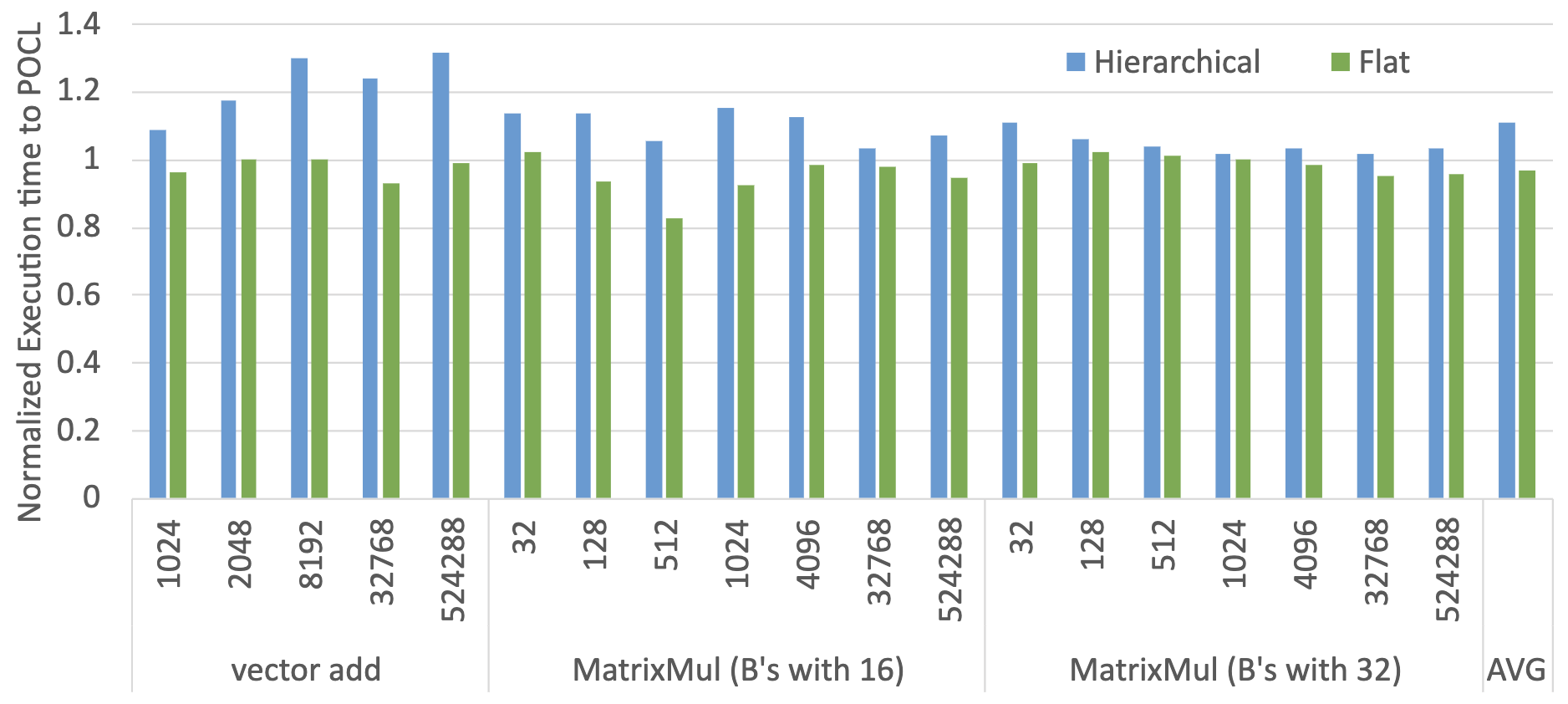}
    \caption{Performance comparisons of flat collapsing and hierarchical collapsing}
    \label{fig:single_nested}
\end{figure}

Although hierarchical collapsing can support wrap-level features, it generates nested loops instead of a single loop as flat collapsing does. The complex nested loops incurs more instructions and also makes it difficult to do some optimizations. Figure~\ref{fig:single_nested} shows the overhead of hierarchical collapsing over flat collapsing on X64 architecture, for three micro-benchmarks by varying the vector/matrix sizes, and none of these three benchmarks use warp-level functions. As the results show, hierarchical collapsing downgrades performance by 13\% on average due to additional instructions. Hence, COX uses hybrid-mode: for each input kernel, first checkes whether there are warp-level functions or other features for which cannot be supported by flat collapsing. If not, flat collapsing is used in default.  

\subsubsection{Normal mode vs JIT mode \label{sec:jit_mode}}

\begin{figure}[htbp]
    \centering
    \includegraphics[width=0.7\columnwidth]{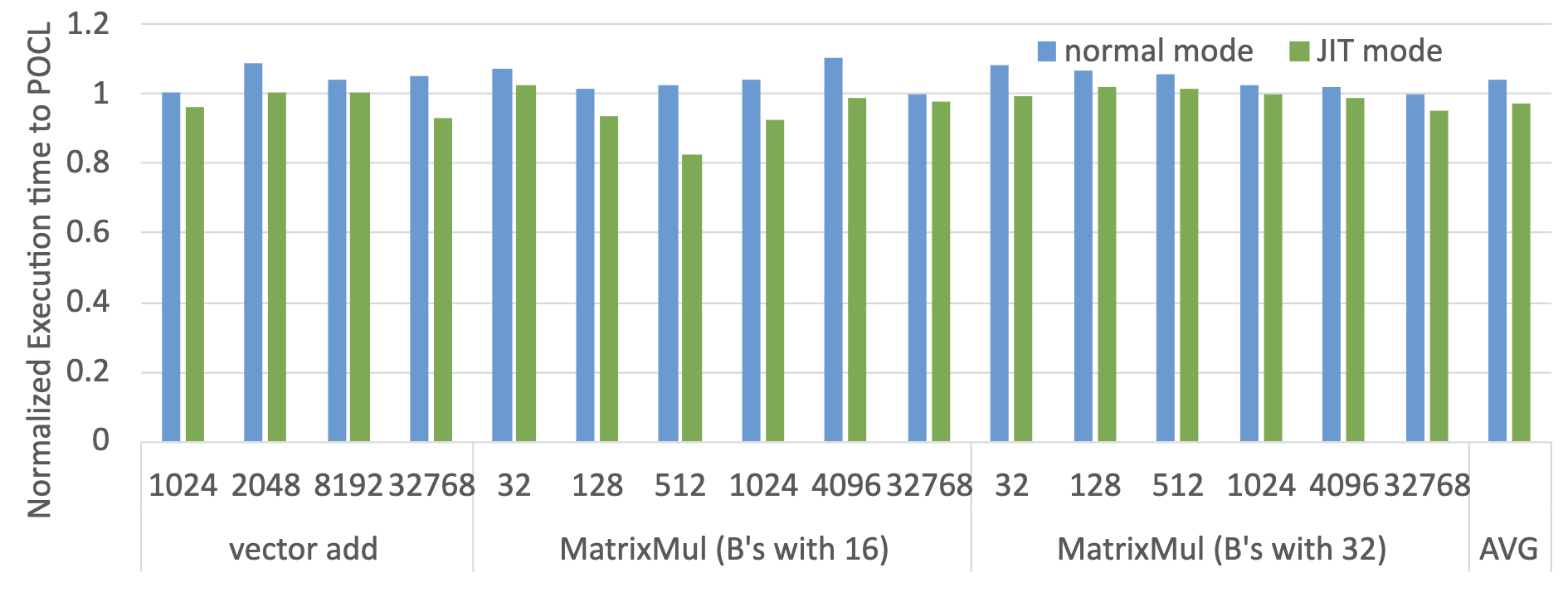}
    \caption{JIT mode can generate faster programs, especially for complex kernels.}
    \label{fig:two_mode}
\end{figure}

Loop optimization is an important optimizations for high-performance programs. Although the intra-warp loop's length is always 32, the inter-warp loop's length depends on the block size, a runtime configuration. COX supports two compile modes: normal mode and JIT mode. Although the programs generated by these two modes will both forward to LLVM's optimizer (with $-O3$ flag), they have an obvious difference, especially when compiling complex kernels. Figure~\ref{fig:two_mode} shows the difference in execution time between two modes. These two modes have a relatively small difference for the VectorAdd kernel, as it is quite simple and can easily be vectorized with compiler optimization even the block size is not provided at compile time. However, for more complex kernels, JIT mode generates programs with higher performance.

\subsubsection{SIMD instructions \label{sec:simd}}
For CPU programs, SIMD instructions are necessary to achieve high performance \cite{jeong2012performance,agulleiro2015tomo3d,bramas2017fast,pharr2012spmd}. The warp vote function execution time with/without AVX is shown in Table~\ref{table:simd_warp_vote}. 
With AVX instructions, around 10x speedup is achieved for both functions. The benefit is due to fewer instructions and branches. 

\begin{table}[]
\small
\centering
\begin{tabular}{|c|c|c|c|}
\hline
function                  &       & w/ AVX        & w/o AVX           \\ \hline
\multirow{5}{*}{vote any} & time (µs)     & 0.241         & 2.542             \\ \cline{2-4} 
                          & instructions  & 1,447,901,852 & 23,472,339,251    \\ \cline{2-4} 
                          & branches      & 100,110,593   & 4,260,452,162     \\  \hline 
\multirow{5}{*}{vote all} & time (µs)     & 0.236         & 2.992             \\ \cline{2-4} 
                          & instructions  & 1,384,476,021 & 29,552,745,486    \\ \cline{2-4} 
                          & branches      & 100,108,177   & 5,220,517,219     \\
                          \hline 
\end{tabular}
\caption{Both functions gain around 10x speed up when using AVX instructions.}
\label{table:simd_warp_vote}
\end{table}

\subsubsection{Scalability}
\begin{figure}[htbp]
    \centering
    \includegraphics[width=90mm]{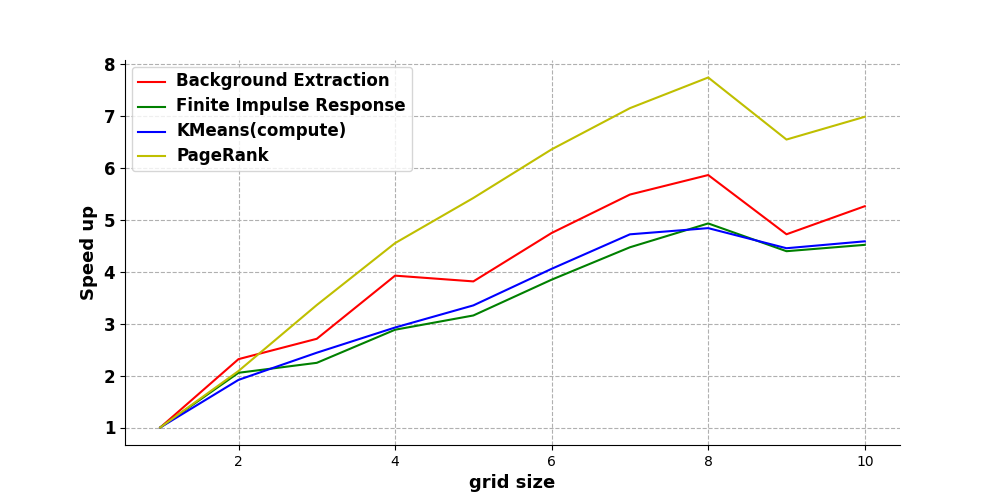}
    \caption{Multi-core execution time with COX.}
    \label{fig:scale}
  \end{figure}

Besides the single block execution time, the execution time for multi-block cases are also measured and the result is shown in Figure~\ref{fig:scale}. In the Hetero-mark benchmark, the kernels have fixed block sizes. Thus, to enlarge the grid size, workload size is also enlarged. As the X86 platform has eight CPU cores; the speed up significantly degrades when the grid sizes are larger than eight. Up to eight, it shows good scalability. 


\section{Related Work \label{sec:related_work}}
The CPU architecture belongs to MPMD, while the GPU architecture is SPMD. Although users can naively execute a GPU thread with a CPU thread, due to the limited parallelism in CPU, the system can only execute around 100 CPU threads simultaneously, which is much smaller than the parallelism in the GPU architecture. Thus, to achieve the same number of threads as a GPU, the CPU has to create more threads than it can actually execute simultaneously, which will incur a large amount of thread context switching overhead. Two methods solve this issue. The first is to accelerate the context-switching time in the CPU. Some researchers extend the CPU architecture to accelerate the context switching~\cite{chen2018enabling}, these hardware-level extensions are beyond the scope of this paper. In the software level, \cite{gummaraju2010twin} proposes to use lightweight threading to accelerate the context switching. Context switching only stores and reloads a few registers, while maintaining the stack memory. Most modifications are in the runtime level, and users can directly use the original GPU source code. As reported in \cite{stratton2013performance}, the AMD CPU OpenCL implementation is based on this technology. However, even with these optimizations, there is still significant overhead for context switching, around 10ns per switching. \\
Thus, another direction is being explored: increasing the workload of each CPU thread. For each CPU thread, instead of executing a single GPU thread, it executes whole GPU threads within a block. This mechanism can elicit two benefits: 1) it can increase the CPU execution time to make it much larger  so that context switching overload becomes negligible; 2) with more workload in a single thread, there are more opportunities to do optimizations (e.g., vectorization, loop transformation). This mechanism has several different names: microthreading~\cite{stratton2010efficient}, thread aggregation~\cite{zhang2013improving}, thread-fusion~\cite{diamos2010ocelot}, region-based serialization~\cite{stratton2013performance}, loop chunking~\cite{shirako2009chunking}, and kernel serialization~\cite{blomkvist2021cumulus}. In this paper, this mechanism is given a new name: flat collapsing. In \cite{stratton2008mcuda, stratton2010efficient}, the authors propose to wrap an SPMD kernel with a loop, and the loop size is equal to the block size. Thus, each loop iteration can simulate a GPU thread within a block, and the a CPU thread is mapped to a GPU block. An important technical detail is supporting synchronization instructions: compilers should separately wrap instructions before/after a synchronization instruction into different loops to maintain the correctness. A similar technology is also discussed in \cite{shirako2009chunking} which utilizes loop transformations (e.g., loop strip-mining, interchange, distribution, unswitching) to transform SPMD execution models with synchronization to Task Parallel execution models. The authors in \cite{zhang2013improving} propose improved static analysis to vectorize the generated loop-programs to improve the performance, and also propose another algorithm to wrap the original kernels with loops to avoid additional synchronization points in previous works. In some GPU architectures, such as NVIDIA GPU, there is implicit lock step within a group of threads. \cite{guo2011correctly} proposed transformations to detect these implicit warp-level synchronizations and maintained them during transformations. The authors in \cite{stratton2013performance} propose to use C Extensions for Array Notation to further accelerate the generated CPU programs with SIMD execution and better spatial locality. \\
Several projects have been proposed to execute CUDA on non-NVIDIA devices. In the early days, NVIDIA provided an emulation framework~\cite{CudaEmulator} to execute CUDA on a CPU; each thread within a GPU block is executed by a CPU thread. Horus~\cite{elhelw2020horus} is another emulator. It supports parsing and executing NVIDIA PTX instructions on CPU devices. These emulators are for debugging rather than for performance. In MCUDA\cite{stratton2008mcuda}, the authors also use a source-to-source translation to translate CUDA to C with the flat collapsing mechanism. Ocelot~\cite{diamos2010ocelot} uses the same mechanism, but instead of source-to-source translation, it converts in the PTX level to avoid recompiling. MapCG~\cite{hong2010mapcg} is a hybrid computing framework. It uses source-to-source translation to translate CUDA kernels to C programs, which can be executed on a CPU. \cite{lee2014boosting} proposes another framework for hybird-computing based on Ocelot to translate GPU programs on the PTX level. Cumuls~\cite{blomkvist2021cumulus} uses Clang to parse the CUDA programs and modifies them on AST level. Cumuls is mainly concern on CUDA runtime support, as for compilation part, it reuses the transformation in MCUDA. Instead of directly translating CUDA/PTX to CPU executable files, other projects utilize the portability of other front-end languages. The authors in \cite{harvey2011swan, sathre2019portability} propose using source-to-source translation to translate CUDA  to OpenCL. Instead of source-to-source translation, \cite{perkins2017cuda,han2021supporting} implements the translations with LLVM IR. DPC++ Compatibility Tool~\cite{dpct} and HIPIFY~\cite{hipify} are frameworks that translate CUDA to source languages for Intel and AMD devices. \\
Most related works only focus on supporting old CUDA features. However, the rapid evolution of GPU hardware and software stacks bring lots of new features which are important to achieve high performance, such as warp-level collectives, unified memory and CudaGraph. Achieving high coverage on these new features is still an ongoing project. The researchers in \cite{patel2021virtual} propose to use explicitly barriers and memory exchanges to support warp shuffle on OpenMP, which shares the same insight with COX. \\
OpenCL is another framework that supports executing SPMD programs on MPMD architectures. POCL~\cite{jaaskelainen2015pocl} is an open source OpenCL implementation which supports CPU backend. To support SPMD programs on CPU, POCL implements flat collapsing on LLVM IR level. The authors in ~\cite{karrenberg2012improving} also proposes to use flat collapsing on OpenCL, but with a different method to insert extra synchronization and find Parallel Region which result in fewer extra synchronization barriers. However, this method is not extendable for Hierarchical Parallel Region, thus, cannot be utilized to support warp-level features. In \cite{kim2012snucl}, another OpenCL implementation has been proposed, which mainly focus on support OpenCL programs on multi-device clusters with heterogeneous devices.
\ignore{
Table~\ref{table:cuda_sdk} summarizes the capability of other frames that support executing CUDA (or OpenCL) on x86 platforms. All projects \cite{stratton2008mcuda,diamos2010ocelot,stratton2010efficient} use flat collapsing ~\cite{kar:hac11} without considering warp-level functions and warp-level barrier. \cite{guo2011correctly} proposed transformations to detect these implicit warp-level synchronization and maintained them during transformations. \\
We mainly focused on CUDA, but there are also some projects that focus on supporting other GPU programs on CPU. For example, \cite{jaaskelainen2015pocl} is an open-source OpenCL implementation. It can support executing OpenCL's SPMD kernels on CPU, using the similar transformations proposed by \cite{stratton2008mcuda}, but has the same limitations of flat collapsing. 
\cite{lomont2011introduction, raman2000implementing} proposed to utilize SIMD instructions on the generated kernels. However, for some CUDA features, it cannot be converted to a format that can be automatically vectorized. In our framework, we use a built-in library to invoke SIMD instructions into these functions. \cite{lee2014boosting} is a project based on Ocelot
}

\section{Conclusion\label{sec:conclusion}}
This project proposes and builds COX, a framework that supports executing CUDA kernels on CPU devices. It also proposes hierarchical collapsing which can be used to transform SPMD programs to MPMD friendly programs and it supports the latest warp-level functions in CUDA. Using CUDA 10.1 Sample as a benchmark, the previous projects can only support 21 of 31 kernels (coverage=68\%), while COX can support 28 (coverage=90\%). The kernel performance is also compared on X86 and AArch64 architectures and shows that the performance is comparable. COX is based on compile-time transformation. Future work will provide a runtime system to support other CUDA features. 

\newpage
\bibliographystyle{unsrt}  
\bibliography{reference}

\appendix
\section{Proof of the generation kernel}
In this section, we prove that after inserting intra-wap loops and inter-warp loops, each instruction from the input GPU kernel is executed {\tt b\_size} times ({\tt b\_size} is the block size), except the instructions used for loop peeling. 

\textbf{Proof1: For if-body/for-body that contains a warp/block barrier, their conditional branch instructions do not belong to any warp/block-level PR} \\
For if-then construct after Step 3, we insert barriers in the end of if-head (barrier in {\tt if.head}), the beginning of if-exit (barrier in {\tt if.exit}) and the end of if-body (barrier in {\tt if.body2}), shown in Figure \ref{fig:barrier_insert}(a). As for for-latch, we insert two barriers around the conditional branch instruction, as shown in Figure. \ref{fig:barrier_insert}(b). For if.cond, it can only be included in PR together with {\tt if.body1} or {\tt if.exit}. For {\tt if.exit}, it has two precessors, thus we will ignore it line10 in Alg. \ref{alg:intra_warp_PR}. As for {\tt if.body1}, as it does not post-dominate {\tt if.cond}, it will not include {\tt if.cond} into the PR. The proof of {\tt for.cond} is quite same. \\
\textbf{Proof2: All other blocks must belong to one and only one warp/block-level PR} \\
This paragraph only proof the block-level PR. The warp-level PR is quite similar. First, let us prove a block can only belong to at most one block-level PR. If a block {\tt B} belongs to more than one PRs, this block must be the ancestor of two blocks that both have block-level barriers. Assuming the block {\tt NCA} is the nearest common ancestor of these two blocks in the control flow graph, it is obvious that {\tt NCA} is the head of an if-then-construct or the latch of a for-construct. However, from the {\tt Proof1}, we know neither of them belongs to any PR.
In other situations, if a block {\tt B} doesn't belong to any PR, it means all nodes that post-dominates {\tt B} do not have a block-level barrier. However, in Step2, we insert a block-level barrier at the end of the exit block. And exit block will always post-dominates {\tt B}.

\end{document}

%% file: cmd.tex
\usepackage{booktabs}   
\usepackage{subcaption} 
                        
\usepackage[utf8]{inputenc} 
\usepackage[T1]{fontenc}    
\usepackage{hyperref}       
\usepackage{url}            
\usepackage{booktabs}       
\usepackage{amsfonts}       
\usepackage{microtype}      
\usepackage{xcolor}
\usepackage{float} 
\usepackage{fancyhdr}

\usepackage{graphicx}
\usepackage{multirow}
\usepackage{cite}
\usepackage{appendix}
\usepackage{amsmath}

\usepackage{listings}

\usepackage{algorithm}
\usepackage{algorithmicx}
\usepackage{algpseudocode}
\MakeRobust{\Call}
\floatname{algorithm}{Algorithm}

\algnewcommand{\LeftComment}[1]{\Statex \(\triangleright\) #1}

\lstdefinestyle{mystyle}{
    frame=tb,
    commentstyle=\color{codegreen},
    keywordstyle=\color{magenta},
    numberstyle=\tiny\color{codegray},
    stringstyle=\color{codepurple},
    basicstyle={\scriptsize\ttfamily},
    breakatwhitespace=false,         
    breaklines=true,                 
    captionpos=b,                    
    keepspaces=true,                 
    numbers=left,                    
    numbersep=5pt,                  
    showspaces=false,                
    showstringspaces=false,
    showtabs=false,                  
    tabsize=2,
    xleftmargin=1ex,
    belowskip=-0.5em,
    aboveskip=0.5em,
}
\usepackage{amssymb}
\usepackage{pifont}
\newcommand{\cmark}{\ding{51}}%
\newcommand{\xmark}{\ding{55}}%

\definecolor{codegreen}{rgb}{0,0.6,0}
\definecolor{codegray}{rgb}{0.5,0.5,0.5}
\definecolor{codepurple}{rgb}{0.58,0,0.82}
\definecolor{backcolour}{rgb}{0.95,0.95,0.92}
\definecolor{codered}{rgb}{1.0,0.01,0.24}
\definecolor{codeblue}{rgb}{0.0, 0.18, 0.65}

\newcommand{\ignore}[1]{}

\newcommand{\greenv}{\textcolor{codegreen}{\cmark}}
\newcommand{\redx}{\textcolor{codered}{\xmark}}

\usepackage{array}
\newcommand{\PreserveBackslash}[1]{\let\temp=\\#1\let\\=\temp}
\newcolumntype{C}[1]{>{\PreserveBackslash\centering}p{#1}}
\newcolumntype{R}[1]{>{\PreserveBackslash\raggedleft}p{#1}}
\newcolumntype{L}[1]{>{\PreserveBackslash\raggedright}p{#1}}